\documentclass[conference]{IEEEtran}

\usepackage[total={7.0in,9.5in}, top=0.75in, bottom=0.75in, left=0.75in, right=0.75in]{geometry}
\usepackage[cmex10]{amsmath}  
\usepackage{amsfonts}
\usepackage{amssymb}
\usepackage{amsthm}  
\usepackage{algorithm,algorithmic}
\usepackage{graphicx}
\usepackage{psfrag}
\usepackage[caption=false,font=footnotesize]{subfig}  

\newcommand{\integers}{{\mathbb{Z}}}
\newcommand{\connects}{\sim}  

\newtheorem{theorem}{Theorem}
\newtheorem{lemma}[theorem]{Lemma}
\newtheorem{definition}{Definition}
\newtheorem{example}{Example}

\begin{document}

\title{Scalable Constructions of Fractional Repetition Codes in
Distributed Storage Systems}

\author{\IEEEauthorblockN{Joseph~C.~Koo and John~T.~Gill,~III}
\IEEEauthorblockA{Department of Electrical Engineering, Stanford University\\
Email: \texttt{\{jckoo,gill\}@stanford.edu}}
}

\maketitle

\bstctlcite{IEEEexample:BSTcontrol}  

\begin{abstract}
In distributed storage systems built using commodity hardware, it is
necessary to have data redundancy in order to ensure system
reliability.  In such systems, it is also often desirable to be able
to quickly repair storage nodes that fail.  We consider a
scheme---introduced by El Rouayheb and Ramchandran---which uses
combinatorial block design in order to design storage systems that
enable efficient (and exact) node repair.  In this work, we
investigate systems where node sizes may be much larger than
replication degrees, and explicitly provide algorithms for
constructing these storage designs.  Our designs, which are related to
projective geometries, are based on the construction of bipartite cage
graphs (with girth~$6$) and the concept of mutually-orthogonal Latin
squares.  Via these constructions, we can guarantee that the resulting
designs require the fewest number of storage nodes for the given
parameters, and can further show that these systems can be easily
expanded without need for frequent reconfiguration.
\end{abstract}


\section{Introduction}
\label{sec:intro}

Recent trends in distributed storage systems have been toward the use
of commodity hardware as storage nodes, where nodes may be
individually unreliable.  Such systems can still be feasible for
large-scale storage as long as there is overall reliability of the
entire storage system.  Recent research in distributed storage systems
has focused on using techniques from coding theory to increase storage
efficiency, without sacrificing system reliability and node
repairability~\cite{dimakis:netcoding_dist_storage}.

In this work, we consider storage systems where failed storage nodes
must be quickly replaced by replacement nodes.  To achieve short
downtimes, we consider techniques where the repair of a particular
node (i.e., by obtaining replacement data) is via contacting multiple
non-failed nodes in parallel---where each contacted node contributes
only a small portion of the replacement data.  Such replacement
strategies have been studied in the context of both \emph{functional
repair}~\cite{dimakis:netcoding_dist_storage}---where replacement
nodes serve functionally for overall data recovery---and \emph{exact
repair}---where replacement nodes must be exact copies of the failed
node.

We build upon the work of El Rouayheb and
Ramchandran~\cite{elrouayheb:frac_rep_codes}, who propose a storage
system allowing for exact repair.  Using the idea of Steiner
systems~\cite{cameron:combinatorics}, the authors design distributed
storage systems with the desired redundancy and repairability
properties---where even though each storage node is responsible for
storing multiple data chunks, replacement of any failed node is always
possible by obtaining only a single data chunk from each of several
non-failed nodes.  In systems where multiple nodes can be read in
parallel, then such a scheme ensures high availability, even in the
presence of node failures.  Moreover, since the scheme described in
\cite{elrouayheb:frac_rep_codes} stores data in an uncoded manner, for
computing applications the storage nodes may also serve as processing
nodes.

A Steiner system $S(t,k,v)$ specifies a distribution of $v$ elements
into blocks of size~$k$ such that the maximum number of overlapping
elements between any two blocks is $t - 1$ (so if $t = 2$, then no two
blocks can share any pairs of elements\footnotemark).
For instance, Example~\ref{ex:steiner_3_9} shows a Steiner system and
the resulting distribution of data chunks to storage nodes.%
\footnotetext{In the rest of this paper, whenever we use the term
Steiner system, we are referring to Steiner systems with $t = 2$.}

\begin{figure}[htbp]
\centering
\newcommand{\bsz}{\small}  
\newcommand{\gsz}{\small}  
\psfrag{b0}[c][c]{\bsz $b_{0}$}  \psfrag{b1}[c][c]{\bsz $b_{1}$}
\psfrag{b2}[c][c]{\bsz $b_{2}$}  \psfrag{b3}[c][c]{\bsz $b_{3}$}
\psfrag{b4}[c][c]{\bsz $b_{4}$}  \psfrag{b5}[c][c]{\bsz $b_{5}$}
\psfrag{b6}[c][c]{\bsz $b_{6}$}  \psfrag{b7}[c][c]{\bsz $b_{7}$}
\psfrag{b8}[c][c]{\bsz $b_{8}$}  \psfrag{b9}[c][c]{\bsz $b_{9}$}
\psfrag{bA}[c][c]{\bsz $b_{10}$} \psfrag{bB}[c][c]{\bsz $b_{11}$}
\psfrag{g00}[c][c]{\gsz $0$}  \psfrag{g01}[c][c]{\gsz $1$}  \psfrag{g02}[c][c]{\gsz $2$}
\psfrag{g10}[c][c]{\gsz $0$}  \psfrag{g11}[c][c]{\gsz $3$}  \psfrag{g12}[c][c]{\gsz $6$}
\psfrag{g20}[c][c]{\gsz $0$}  \psfrag{g21}[c][c]{\gsz $4$}  \psfrag{g22}[c][c]{\gsz $8$}
\psfrag{g30}[c][c]{\gsz $0$}  \psfrag{g31}[c][c]{\gsz $5$}  \psfrag{g32}[c][c]{\gsz $7$}
\psfrag{g40}[c][c]{\gsz $1$}  \psfrag{g41}[c][c]{\gsz $3$}  \psfrag{g42}[c][c]{\gsz $8$}
\psfrag{g50}[c][c]{\gsz $1$}  \psfrag{g51}[c][c]{\gsz $4$}  \psfrag{g52}[c][c]{\gsz $7$}
\psfrag{g60}[c][c]{\gsz $1$}  \psfrag{g61}[c][c]{\gsz $5$}  \psfrag{g62}[c][c]{\gsz $6$}
\psfrag{g70}[c][c]{\gsz $2$}  \psfrag{g71}[c][c]{\gsz $3$}  \psfrag{g72}[c][c]{\gsz $7$}
\psfrag{g80}[c][c]{\gsz $2$}  \psfrag{g81}[c][c]{\gsz $4$}  \psfrag{g82}[c][c]{\gsz $6$}
\psfrag{g90}[c][c]{\gsz $2$}  \psfrag{g91}[c][c]{\gsz $5$}  \psfrag{g92}[c][c]{\gsz $8$}
\psfrag{gA0}[c][c]{\gsz $3$}  \psfrag{gA1}[c][c]{\gsz $4$}  \psfrag{gA2}[c][c]{\gsz $5$}
\psfrag{gB0}[c][c]{\gsz $6$}  \psfrag{gB1}[c][c]{\gsz $7$}  \psfrag{gB2}[c][c]{\gsz $8$}
\includegraphics[height=1.1in,scale=0.5]{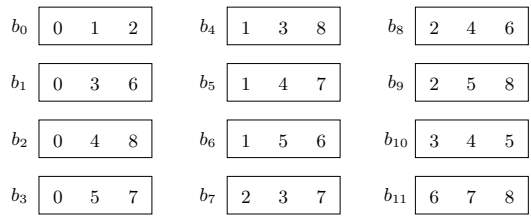}
\caption{Storage design from Steiner system $S(2,3,9)$; same as
\cite[Fig.~6(a)]{elrouayheb:frac_rep_codes}.}
\label{fig:steiner_3_9}
\end{figure}

\begin{example}
\label{ex:steiner_3_9}
Consider a distributed storage system to store $9$~total data chunks,
where each chunk is stored within storage nodes that can hold
$3$~chunks each.  Then it is possible to distribute the chunks across
$12$~nodes, where every chunk has exactly $4$~replicas and any two
distinct nodes share at most only one overlapping chunk.  This is
shown in Figure~\ref{fig:steiner_3_9}.
\end{example}

In most practical distributed storage systems, however, it is often
desirable for the number of data chunks per node\footnotemark ~to be
much greater than the replication degree of each chunk.  For example,
the Google File System~\cite{ghemawat:google_file_system}---which
stores data in chunks of as small as 64~MB each---has a replication
degree on the order of three replicas but may store thousands of
chunks on each storage node.  Thus in this work, we consider a
graph-based construction of Steiner systems where the replication
degree and node size are significantly asymmetric.%
\footnotetext{For brevity, we refer to the number of chunks per node
as the \emph{node size}.}

Specifically, we construct storage systems where the replication
degree of each data chunk is $q+1$, whereas each node may
store up to $q^n + q^{n-1} + \cdots + q^2 + q + 1$ chunks (for any
given integer~$n$).  Although it is known from the theory of
projective geometries~\cite{cameron:combinatorics} that systems with
these parameters can be designed, by using our graph-based method we
are able to give a systematic construction that is highly scalable;
for a system constructed according to the methods in this paper, it is
always possible to increase the storage system without moving any
existing data chunks---and still be able to preserve the property that
no pairs of chunks recur in more than one storage node.

Our construction is based on relating Steiner system problems with the
problem of shortest cycles on bipartite graphs.  More specifically,
our systems arise from the construction of \emph{cage
graphs}~\cite{wong:cages}, which are graphs with the minimum number of
vertices for a given allowable shortest cycle length and other
specified conditions on the vertex degrees.  Because we are
constructing cage graphs, we further know that for a given desired
node size and replication degree, our constructions are the
\emph{smallest} possible systems (in terms of total number of storage
nodes and total number of data chunks stored).  This is useful for the
practical application of such constructions, as it immediately
translates into least hardware cost for the desired system
requirements.

\subsection{Related Work}
\label{subsec:intro_related}

The problem of distributed storage with efficient repair is discussed
in \cite{dimakis:netcoding_dist_storage}.  Using network coding, the
authors propose a scheme for storing data where node repair is
functional.  Dimakis et al.~\cite{dimakis:netcoding_dist_storage} also
define the idea of a storage-bandwidth tradeoff, and discuss ways to
implement either minimum storage or minimum bandwidth systems.  Even
though exact repair of storage nodes is sometimes necessary, the
storage-bandwidth tradeoff under exact repair is not yet fully
understood.  Building upon the network coding constructions of
\cite{dimakis:netcoding_dist_storage}, Rashmi et
al.~\cite{rashmi:explicit_opt_exact_regen_codes_min_bandwidth} give a
scheme for achieving the minimum bandwidth operating point under exact
repair, finding a point on the storage-bandwidth tradeoff curve.

El Rouayheb and Ramchandran~\cite{elrouayheb:frac_rep_codes} introduce
a related scheme, termed \emph{fractional repetition codes}, which can
perform exact repair for the minimum bandwidth regime.  They then
derive information theoretic bounds on the storage capacity of such
systems with the given repair requirements.  Although their repair
model is table-based (instead of random access as in
\cite{dimakis:netcoding_dist_storage}), the scheme of
\cite{elrouayheb:frac_rep_codes} has the favorable characteristics of
exact repair and the uncoded storage of data chunks.  Randomized
constructions of such schemes are investigated in
\cite{pawar:dress_codes_randomized}.

Uncoded storage has numerous advantages for distributed storage
systems.  For instance, uncoded data at nodes allows for distributed
computing (e.g., for cloud computing), by spreading out computation to
the node(s) that contain the data to be processed.  Upfal and
Widgerson~\cite{upfal:share_memory_dist_sys} consider a method for
parallel computation by randomly distributing data chunks among
multiple memory devices, and derive some asymptotic performance
results.  In contrast, our designs are deterministic, and we are also
able to guarantee the smallest possible size for our storage system.
Furthermore, if uncoded data chunks are distributed among the nodes
according to Steiner systems, then load-balancing of computations is
always possible.

Steiner systems are an example of balanced incomplete block design
(BIBD), within the field of combinatorial design
theory~\cite{lindner:design_theory}.  Some parameters for which
Steiner systems can be designed are given in
\cite{colbourn:handbook_comb_designs, elrouayheb:frac_rep_codes}.  In
this work, we consider Steiner systems similar to those from finite
projective planes.  Specifically, designs in which the replication
degree is $q+1$ and with each storage node storing up to $q^n +
q^{n-1} + \cdots + q^2 + q + 1$ data chunks can also be found from the
projective geometry $PG(n+1,q)$---where the data chunks are the lines
and the storage nodes are the points of the corresponding space.
However, in this work we show that via our recursive graph
construction method, it is possible to initially deploy small storage
systems without needing to know \emph{a priori} the future maximum
extent of the storage system---while still being able to preserve the
Steiner property in subsequent expanded systems.\footnotemark  ~This
alternate approach for constructing projective geometries has
tremendous benefits for practical storage system designs, as otherwise
the connection between system design and the construction and
extension of such geometries is not immediately obvious.  Furthermore,
our graph-based construction is simple to implement, and designs are
uniquely determined given knowledge of the base set of
mutually-orthogonal Latin squares (which we discuss later).%
\footnotetext{We do not describe this in detail, but very similar
graph-based methods can also be used to construct designs related to
the affine geometry $AG(n+1,q)$.  These constructions are just as
expandable as the projective geometry--based designs.  A brief note on
these constructions is given in
Section~\ref{subsec:scal_designs_other}.}

In addition to \cite{elrouayheb:frac_rep_codes}, the use of BIBDs for
guaranteeing load-balanced disk repair in distributed storage systems
is also considered in \cite{muntz:perf_disk_array_under_failure,
holland:archs_algs_failure_recovery}, for application to RAID-based
disk arrays.  In \cite{holland:archs_algs_failure_recovery}, the
authors discuss how block designs may be used to lay out parity
stripes in declustered parity RAID disk arrays.  The block designs
from our work may be helpful for distributing parity blocks in this
scenario, in order to build disk arrays with good repair properties.

Certain block designs may also be applicable to the design of
error-correcting codes, particularly in the construction of
geometrical codes~\cite[Sections 2.5 and
13.8]{macwilliams:error_corr_codes}.  Graphs without short cycles have
been considered in the context of Tanner
graphs~\cite{tanner:recursive_approach_low_complex_codes}, and finite
geometries in particular have been considered in the context of LDPC
codes~\cite{kou:ldpc_finite_geometries}.  Block designs and their
related bipartite graphs are also considered in code design for
magnetic recording applications in
\cite{vasic:steiner_magnetic_recording}.

\subsection{Outline of Paper}
\label{subsec:intro_outline}

In the next section, we provide necessary background.
Section~\ref{sec:reg_cages} illustrates how our constructions work,
through the construction of regular bipartite cage graphs; this
construction provides a base upon which the larger construction of
Section~\ref{sec:scal_designs} is built.  In
Section~\ref{sec:scal_designs} we give the main contribution, which is
the design of scalable storage systems that can be expanded readily.
Finally, Section~\ref{sec:conclusion} concludes.

\section{Preliminaries}
\label{sec:prelims}

\subsection{Notation}
\label{subsec:prelims_notation}

When describing parameters for constructible graphs we let $p_n(q) =
q^n + q^{n-1} + \cdots + q^2 + q + 1 = \frac{q^{n+1}-1}{q-1}$, for $n
\in \integers_{++}$.  In the rest of this paper, $q$ will always
denote either a prime number or a power of a prime.

In this work, we consider simple undirected bipartite graphs $G =
(X,Y,E)$.   Cardinality is denoted by $|\cdot|$.  For a vertex~$x$,
$\deg(x)$ gives the number of incident edges.  We only consider graphs
where all of the vertices in a vertex set have the same degree, so we
can write $\deg(X) = \deg(x)$ for some $x \in X$.  The symbol
$\connects$ is used to denote an edge; for vertices $x$ and $y$, we
say that $x \connects y$ if and only if $(x,y) \in E$.

\subsection{Graph Interpretation of Steiner Systems}
\label{subsec:prelims_steiner_sys}

A Steiner system is a collection of elements, $\mathcal{V}$, into
blocks, $\mathcal{B}$, where any subset of elements only occurs once
in the block collection.  We reinterpret the Steiner system
requirements by considering its incidence
graph~\cite{bondy:graph_theory}.  In this work, we will consider
bipartite graphs $G = (X,Y,E)$, where there are $u$~vertices in $X$,
each of degree~$k$, and $v$~vertices in $Y$, each of degree~$l$.  We
call such a graph to be \emph{biregular} when $k \ne l$.  Clearly, $lv
= uk$.

\begin{figure}[htbp]
\centering
\psfrag{y00}[l][l]{\scriptsize $0$}
\psfrag{y01}[l][l]{\scriptsize $1$}
\psfrag{y02}[l][l]{\scriptsize $2$}
\psfrag{y03}[l][l]{\scriptsize $3$}
\psfrag{y04}[l][l]{\scriptsize $6$}
\psfrag{y05}[l][l]{\scriptsize $4$}
\psfrag{y06}[l][l]{\scriptsize $8$}
\psfrag{y07}[l][l]{\scriptsize $5$}
\psfrag{y08}[l][l]{\scriptsize $7$}
\psfrag{x00}[c][l]{\tiny $\{0,1,2\}$}
\psfrag{x01}[c][l]{\tiny $\{0,3,6\}$}
\psfrag{x02}[c][l]{\tiny $\{0,4,8\}$}
\psfrag{x03}[c][l]{\tiny $\{0,5,7\}$}
\psfrag{x04}[c][l]{\tiny $\{3,4,5\}$}
\psfrag{x05}[c][l]{\tiny $\{6,8,7\}$}
\psfrag{x06}[c][l]{\tiny $\{1,4,7\}$}
\psfrag{x07}[c][l]{\tiny $\{1,3,8\}$}
\psfrag{x08}[c][l]{\tiny $\{1,6,5\}$}
\psfrag{x09}[c][l]{\tiny $\{2,8,5\}$}
\psfrag{x10}[c][l]{\tiny $\{2,3,7\}$}
\psfrag{x11}[c][l]{\tiny $\{2,6,4\}$}
\psfrag{Y}[l][l]{elements}
\psfrag{X}[l][l]{blocks}
\includegraphics[height=0.62in,scale=0.5]{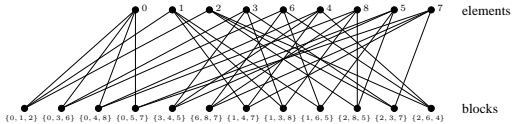}
\caption{Bipartite graph of Steiner system corresponding to $k = 3$, $l = 4$.}
\label{fig:steiner_sys_cage_bireg_k_3_l_4_bipartite}
\end{figure}

Now we label the vertices of $Y$ as the elements of $\mathcal{V}$
(i.e., $\mathcal{V} = \{y_g \; | \; g = 0,1,\ldots,v-1\}$), and the
vertices of $X$ as the blocks of $\mathcal{B}$.  Consider a particular
vertex $x_h$ (where $h \in \{0,1,\ldots,u-1\}$), and define block $b_h
= \{y_g \in Y \; | \; y_g \connects x_h\}$.  Then the collection of
blocks, $\mathcal{B} = \{b_h \; | \; h = 0,1,\ldots,u-1\}$, satisfies
the following:
\begin{enumerate}
\item Each element $y_g \in \mathcal{V}$ occurs in exactly $l$~blocks
of $\mathcal{B}$.
\item Each block $b_h \in \mathcal{B}$ contains $k$~elements.
\end{enumerate}
It is clear that whenever two blocks $b_h$ and $b_{h'}$ share some
pair of elements $y_g$ and $y_{g'}$, then this is equivalent to the
$4$-cycle $x_h \connects y_g \connects x_{h'} \connects y_{g'}
\connects x_h$.  Thus the nonexistence of such $4$-cycles is
equivalent to the nonexistence of shared pairs of elements between
blocks.  In Figure~\ref{fig:steiner_sys_cage_bireg_k_3_l_4_bipartite},
we show the bipartite graph associated with
Example~\ref{ex:steiner_3_9}.

In the above, we construct Steiner systems where $l$ is the repetition
degree, $k$ is the block size, $v$ is the total number of elements,
and $u$ is the total number of blocks.  In the rest of this paper, we
shall always assume that $l \geq k$.\footnotemark%
\footnotetext{We can construct systems where $k > l$ by swapping the
two vertex sets.}

Since $X$ and $Y$ are interchangeable, we could instead let $X$ be the
elements and $Y$ be the blocks of another block system,\footnotemark
~resulting in the \emph{transpose codes} of
\cite{elrouayheb:frac_rep_codes}.  ~Since for practical cases we wish
to construct distributed storage systems where the repetition degree
is smaller than the block size, we will more often employ the
transpose code.  To stay consistent with $l \geq k$, in these cases we
let $k$ be the repetition degree and $l$ be the block size.  Under
this interpretation, $u$ is the number of elements and $v$ is the
total number of blocks.%
\footnotetext{In the language of finite geometries, interchanging the
roles of elements and blocks is the same as interchanging
\emph{points} and \emph{lines}.}

\subsection{Cage Graphs}
\label{subsec:prelims_cage_graphs}

In an undirected graph $G = (V, E)$, a cycle of length~$d$ is a set of
$d$~vertices connected in a closed path.  In the bipartite graph $G =
(X,Y,E)$, such a cycle must necessarily alternate between vertices of
$X$ and vertices of $Y$; thus any cycles must have even length.  The
\emph{girth} of a graph is defined as the length of the shortest cycle
in the graph.

Then, a $d$-\emph{cage} is a girth-$d$ graph with minimum number of
vertices for a particular desired degree
distribution~\cite{wong:cages, bondy:graph_theory}.  The goal of this
work is to construct biregular cages of girth~$6$ (so that no
$4$-cycles are present)---in order to construct the smallest possible
Steiner system with the desired parameters.  Using transpose codes we
can then construct systems requiring the fewest possible storage nodes
(i.e., smallest $v$) and the least number of total distinct chunks
(i.e., smallest $u$), while still having the desired repetition
degree~$k$ and block size~$l$.  Such systems will meet the lower bound
of Lemma~\ref{thm:lower_bounds_num_vertices}.\footnotemark%
\footnotetext{The result of Lemma~\ref{thm:lower_bounds_num_vertices}
is sometimes known as a Moore-type
bound~\cite{milenkovic:shortened_array_codes_large_girth}, although we
note that the bound in \eqref{eq:num_elements_lower_bound} is tighter
than the corresponding bound in
\cite{hoory:size_bipartite_graphs_given_girth} for our case, when $l >
k$.}

\begin{lemma}
\label{thm:lower_bounds_num_vertices}
Consider a simple biregular bipartite graph $(X,Y,E)$ that does not
have any cycle of $4$ or fewer vertices.  If $\deg(X) = k$ and
$\deg(Y) = l$ (where $l \geq k$), then the number of vertices, $v =
|Y|$ and $u = |X|$, has lower bounds
\begin{eqnarray}
v & \geq & 1 + l (k-1)  \label{eq:num_blocks_lower_bound} \\
u & \geq & l + l (l-1) (k-1)/k  \label{eq:num_elements_lower_bound}
\mbox{.}
\end{eqnarray}
\end{lemma}
\vspace{-1.0ex}
\begin{IEEEproof}
We sketch the proof for \eqref{eq:num_blocks_lower_bound} here; a more
detailed proof of \eqref{eq:num_blocks_lower_bound} as well as for
\eqref{eq:num_elements_lower_bound} is provided in
Appendix~\ref{app:proof_thm_lower_bounds_num_vertices}.

One method for constructing the bipartite graph is by starting with a
single vertex $y \in Y$ (called the \emph{layer~$0$} vertex) and
connecting it to $l$~vertices of $X$ (called the \emph{layer~$1$}
vertices).  These vertices of $X$ must be connected to $k-1$
distinct other vertices of $Y$ (the \emph{layer~$2$} vertices).  Note
that any remaining vertices of $X$ (the \emph{layer~$3$} vertices)
would then need to be connected back to the layer~$2$ vertices of $Y$
in such a way as to preserve the nonexistence of $4$-cycles.
\end{IEEEproof}

Any bipartite cage achieving the lower bounds of
Lemma~\ref{thm:lower_bounds_num_vertices} satisfies the Steiner system
property that each pair of elements occurs in \emph{exactly} one
block.  We already know that every pair of elements occurs in at most
one block.  Since $v = 1 + l(k-1)$ and $u = l + l(l-1)(k-1)/k$ also
satisfies $\binom{v}{2} = u \binom{k}{2}$,\footnotemark ~we know that
every pair of elements occurs in at least one block---and therefore
occurs in only one block.%
\footnotetext{The condition $\binom{v}{2} = u \binom{k}{2}$ comes from
the fact that there are a total of $\binom{v}{2}$ pairs of elements,
which should correspond exactly to the sets of $\binom{k}{2}$ pairs of
elements in each of the $u$~blocks.}

The proof of Lemma~\ref{thm:lower_bounds_num_vertices} gives us clues
on how to construct bipartite graphs that achieve the lower
bounds---which must necessarily be cage graphs.  We will show how to
avoid introducing $4$-cycles between the layer~$2$ and layer~$3$
vertices, by considering the use of mutually-orthogonal Latin squares
(see Appendix~\ref{app:latin_squares} or \cite{denes:latin_squares}).
Specifically, in order to construct the cage graphs, we will require
the existence of a set of $q$~mutually-orthogonal $q \times q$
squares, $\{L^{(0)}, L^{(1)}, \ldots, L^{(q-1)}\}$, where $L^{(0)}$ is
a square with every column in natural order, and $L^{(1)}, L^{(2)},
\ldots, L^{(q-1)}$ are mutually-orthogonal Latin squares where each
square has its zeroth column in natural order.  Such a set always
exists when $q$ is a prime or a prime power.  We give an example for
$q = 3$:
\begin{example}
\label{ex:mut_orth_squares_q_3}
A set of $3$ mutually-orthogonal $3 \times 3$ squares is
\[
\scriptsize
\begin{array}{l}
L^{(0)} = \left[ \begin{array}{rrr}
0 & 0 & 0 \\ 1 & 1 & 1 \\ 2 & 2 & 2
\end{array} \right],\;\;
L^{(1)} = \left[ \begin{array}{rrr}
0 & 1 & 2 \\ 1 & 2 & 0 \\ 2 & 0 & 1
\end{array} \right],\;\;
L^{(2)} = \left[ \begin{array}{rrr}
0 & 2 & 1 \\ 1 & 0 & 2 \\ 2 & 1 & 0
\end{array} \right]
\end{array}
\mbox{.}
\]
\end{example}

\section{Regular Cage Graphs}
\label{sec:reg_cages}

We now show how to construct girth-$6$ bipartite cage graphs where the
degrees of both vertex sets are equal.  More specifically, the vertex
degrees will satisfy $\deg(X) = \deg(Y) = q+1$ (i.e., $k = l = q+1$),
where $q$ is any prime or power of a prime.  The resulting graphs will
have $|X| = |Y| = q^2 + q + 1$.

\subsection{Construction of Regular Cage Graph}
\label{subsec:reg_cages_constr}

The construction of regular bipartite cage graphs of girth~$6$ is
inspired from the construction in Wong~\cite{wong:cages}, and is given
in Algorithm~\ref{alg:cage_k_q+1_l_q+1}.  Bipartiteness arises from
the construction.

\begin{algorithm}[htb]
\caption{\footnotesize Construction of bipartite cage when $k = l = q+1$}
\label{alg:cage_k_q+1_l_q+1}
\footnotesize{
\begin{algorithmic}[1]
\STATE \label{it:cage_k_q+1_l_q+1_layer_0}
\textbf{[Layer~$0$]}
Start with a single vertex $y_0 \in Y$.
\STATE \label{it:cage_k_q+1_l_q+1_layer_1}
\textbf{[Layer~$1$]}
Connect $y_0$ to $l = q+1$ vertices of $X$.  Without loss of
generality, call these vertices $x_0, x_1, \ldots, x_{l-1}$.
\STATE \label{it:cage_k_q+1_l_q+1_layer_2}
\textbf{[Layer~$2$]}
For each vertex~$x_j$, $j = 0,1,\ldots,l-1$, connect $x_j$ to $k-1 =
q$ vertices of $Y$.  Let $\hat{y}_{j,m}$, $m = 0,1,\ldots,k-2$, denote
the vertices of this step that are connected to vertex $x_j$.
\STATE \label{it:cage_k_q+1_l_q+1_layer_3}
\textbf{[Layer~$3$]}
Connect each vertex~$\hat{y}_{0,m}$ ($m = 0, 1,\ldots,k-2$) to $l-1 =
q$ distinct vertices of $X$, called $\hat{x}_{m,i}$, $i =
0,1,\ldots,l-2$.  Therefore, $\hat{x}_{m,i} \ne \hat{x}_{m',i'}$
unless $m = m'$ and $i = i'$.  There will be $(k-1)(l-1) = q^2$ such
vertices~$\hat{x}_{m,i}$.
\STATE \label{it:cage_k_q+1_l_q+1_connect_latin_squares}
Consider a vertex~$\hat{x}_{m,i}$, where $m \in \{0,1,\ldots,k-2\}$,
$i \in \{0,1,\ldots,l-2\}$.  Connect $\hat{x}_{m,i}$ to
vertices~$\hat{y}_{j+1,L_{i,j}^{(m)}}$, where $j = 0,1,\ldots,l-2$.
\end{algorithmic}
}
\end{algorithm}

The $q^2+q$ layer~$2$ vertices $\hat{y}_{j,m}$, $j = 0,1,\ldots,q$ and
$m = 0,1,\ldots,q-1$, coincide with the vertices $y_1, y_2, \ldots,
y_{q^2+q}$, and can be mapped using $y_{jq+m+1} = \hat{y}_{j,m}$.
Similarly, the $q^2$ layer~$3$ vertices $\hat{x}_{m,i}$, $m =
0,1,\ldots,q-1$ and $i = 0,1,\ldots,q-1$, coincide with the vertices
$x_{q+1}, x_{q+2}, \ldots, x_{q+q^2}$, and can be mapped using
$x_{q+mq+i+1} = \hat{x}_{m,i}$.

Notice that the resulting graph consists of the layer~$0$ and
layer~$2$ vertices on one side of the graph, connected only to
layer~$1$ and layer~$3$ vertices on the other side.

We first show an example of Algorithm~\ref{alg:cage_k_q+1_l_q+1} with
$k = 4$ and $l = 4$ (so $q = 3$), before proving that this indeed
results in the desired cage graph.  This graph will have $|X| = |Y| =
13$.

\begin{figure}[htbp]
\centering
\subfloat[Conclusion of step~\ref{it:cage_k_q+1_l_q+1_layer_3}.]{
\psfrag{y00}[l][l]{\footnotesize $y_{0}$}
\psfrag{y01}[l][l]{\footnotesize $\hat{y}_{0,0}$}
\psfrag{y02}[l][l]{\footnotesize $\hat{y}_{0,1}$}
\psfrag{y03}[l][l]{\footnotesize $\hat{y}_{0,2}$}
\psfrag{y04}[l][l]{\footnotesize $\hat{y}_{1,0}$}
\psfrag{y05}[l][l]{\footnotesize $\hat{y}_{1,1}$}
\psfrag{y06}[l][l]{\footnotesize $\hat{y}_{1,2}$}
\psfrag{y07}[l][l]{\footnotesize $\hat{y}_{2,0}$}
\psfrag{y08}[l][l]{\footnotesize $\hat{y}_{2,1}$}
\psfrag{y09}[l][l]{\footnotesize $\hat{y}_{2,2}$}
\psfrag{y10}[l][l]{\footnotesize $\hat{y}_{3,0}$}
\psfrag{y11}[l][l]{\footnotesize $\hat{y}_{3,1}$}
\psfrag{y12}[l][l]{\footnotesize $\hat{y}_{3,2}$}
\psfrag{x00}[l][l]{\footnotesize $x_{0}$}
\psfrag{x01}[l][l]{\footnotesize $x_{1}$}
\psfrag{x02}[l][l]{\footnotesize $x_{2}$}
\psfrag{x03}[l][l]{\footnotesize $x_{3}$}
\psfrag{x04}[l][l]{\footnotesize $\hat{x}_{0,0}$}
\psfrag{x05}[l][l]{\footnotesize $\hat{x}_{0,1}$}
\psfrag{x06}[l][l]{\footnotesize $\hat{x}_{0,2}$}
\psfrag{x07}[l][l]{\footnotesize $\hat{x}_{1,0}$}
\psfrag{x08}[l][l]{\footnotesize $\hat{x}_{1,1}$}
\psfrag{x09}[l][l]{\footnotesize $\hat{x}_{1,2}$}
\psfrag{x10}[l][l]{\footnotesize $\hat{x}_{2,0}$}
\psfrag{x11}[l][l]{\footnotesize $\hat{x}_{2,1}$}
\psfrag{x12}[l][l]{\footnotesize $\hat{x}_{2,2}$}
\includegraphics[height=1.15in,scale=0.5]{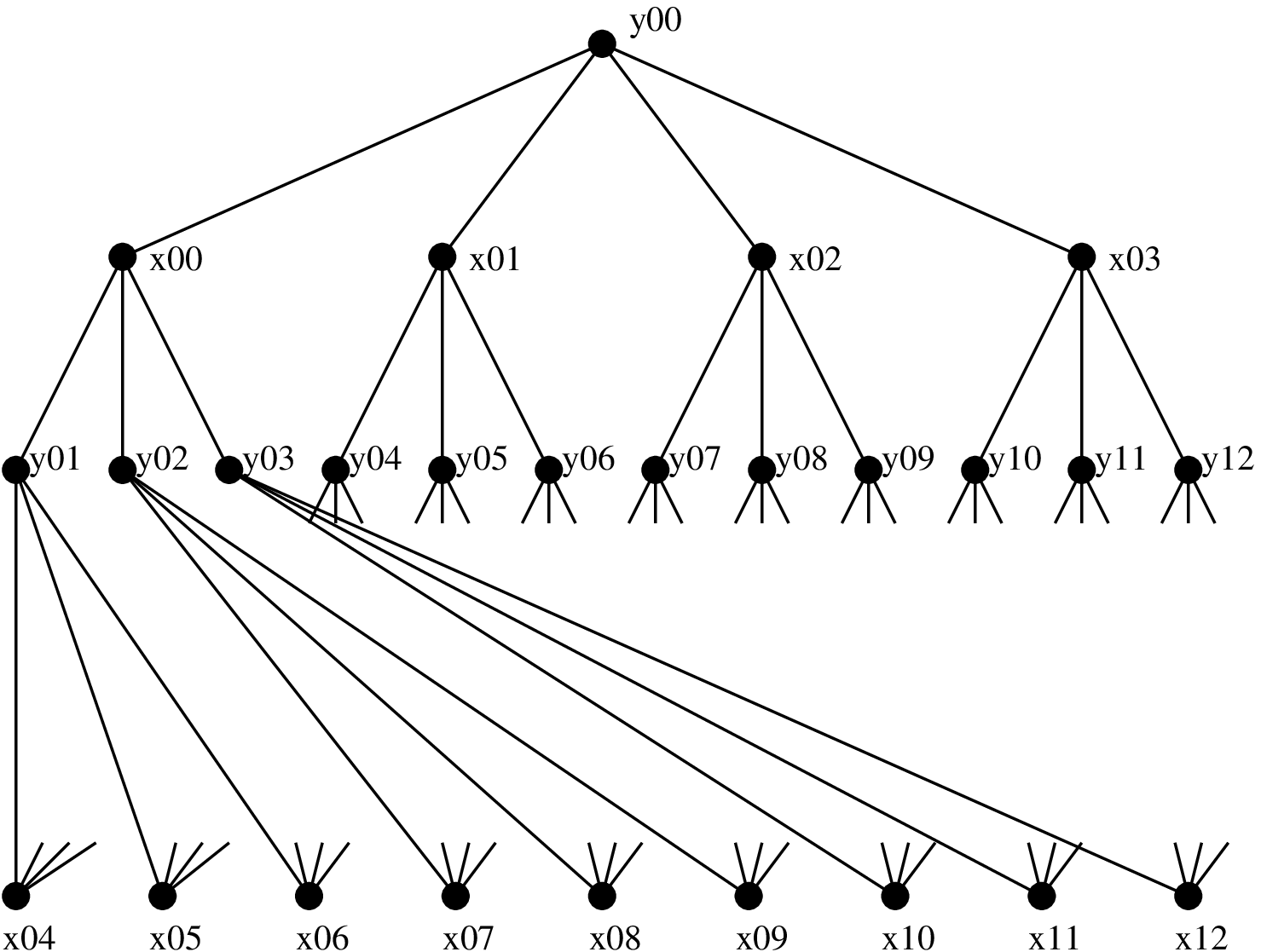}
\label{fig:cage_bireg_k_4_l_4_connect_initial_layer_3}
}
\hfil
\subfloat[Final bipartite graph.]{
\psfrag{y00}[l][l]{\footnotesize $y_{0}$}
\psfrag{y01}[l][l]{\footnotesize $y_{1}$}
\psfrag{y02}[l][l]{\footnotesize $y_{2}$}
\psfrag{y03}[l][l]{\footnotesize $y_{3}$}
\psfrag{y04}[l][l]{\footnotesize $y_{4}$}
\psfrag{y05}[l][l]{\footnotesize $y_{5}$}
\psfrag{y06}[l][l]{\footnotesize $y_{6}$}
\psfrag{y07}[l][l]{\footnotesize $y_{7}$}
\psfrag{y08}[l][l]{\footnotesize $y_{8}$}
\psfrag{y09}[l][l]{\footnotesize $y_{9}$}
\psfrag{y10}[l][l]{\footnotesize $y_{10}$}
\psfrag{y11}[l][l]{\footnotesize $y_{11}$}
\psfrag{y12}[l][l]{\footnotesize $y_{12}$}
\psfrag{x00}[l][l]{\footnotesize $x_{0}$}
\psfrag{x01}[l][l]{\footnotesize $x_{1}$}
\psfrag{x02}[l][l]{\footnotesize $x_{2}$}
\psfrag{x03}[l][l]{\footnotesize $x_{3}$}
\psfrag{x04}[l][l]{\footnotesize $x_{4}$}
\psfrag{x05}[l][l]{\footnotesize $x_{5}$}
\psfrag{x06}[l][l]{\footnotesize $x_{6}$}
\psfrag{x07}[l][l]{\footnotesize $x_{7}$}
\psfrag{x08}[l][l]{\footnotesize $x_{8}$}
\psfrag{x09}[l][l]{\footnotesize $x_{9}$}
\psfrag{x10}[l][l]{\footnotesize $x_{10}$}
\psfrag{x11}[l][l]{\footnotesize $x_{11}$}
\psfrag{x12}[l][l]{\footnotesize $x_{12}$}
\includegraphics[height=1.15in,scale=0.5]{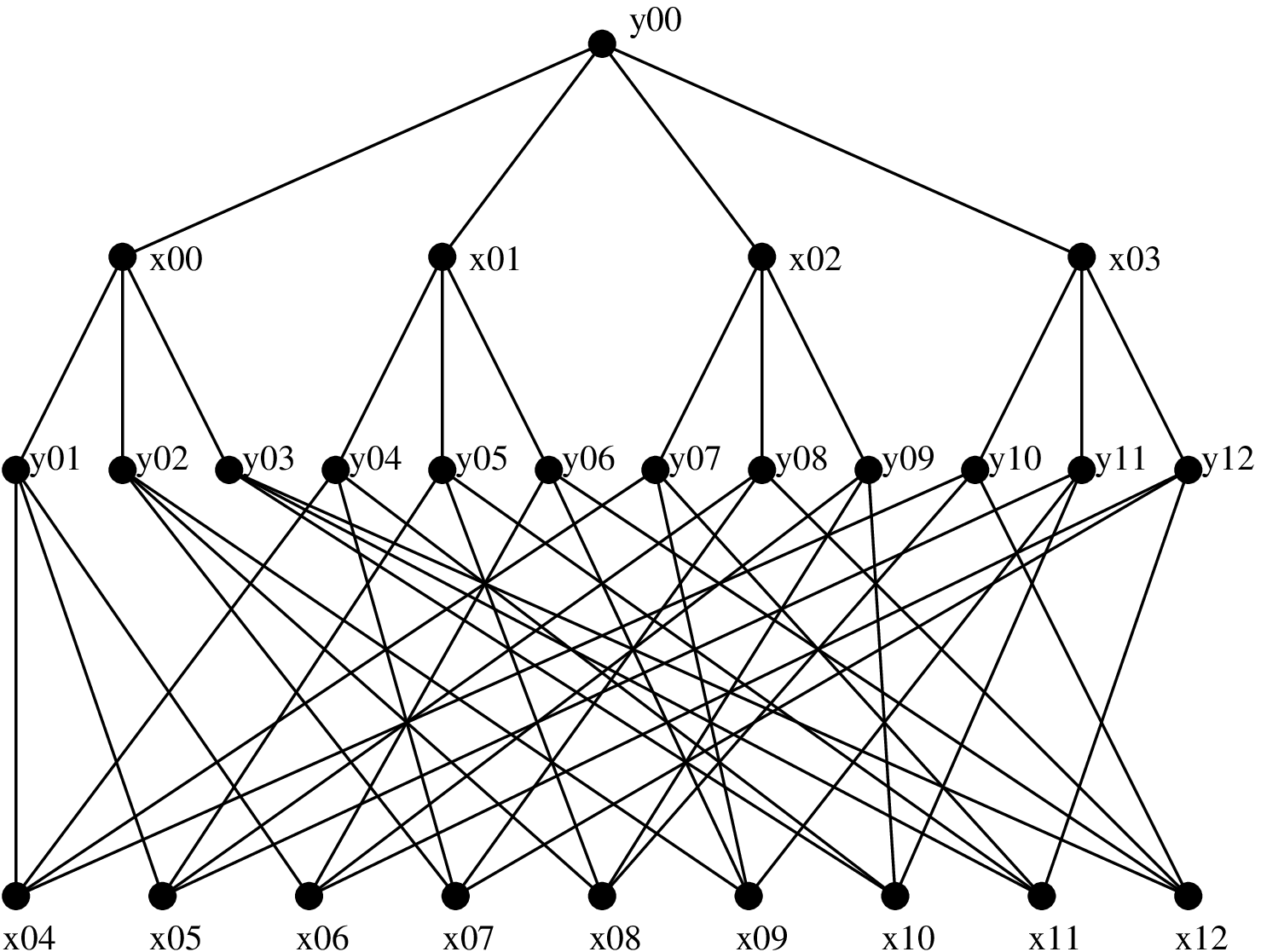}
\label{fig:cage_bireg_k_4_l_4_final}
}
\caption{Construction of bipartite cage where $k = l = 4$, using
Algorithm~\ref{alg:cage_k_q+1_l_q+1}.}
\label{fig:cage_bireg_k_4_l_4}
\end{figure}

The first three steps are straightforward, as they involve connecting
the vertices of layers $0$, $1$, and $2$ in a tree.
Step~\ref{it:cage_k_q+1_l_q+1_layer_3} connects all the vertices
associated with $\hat{y}_{0,m}$ with the $l-1 = q$ vertices
$\hat{x}_{m,i}$, $i = 0,1,\ldots,q-1$.  This gives
Figure~\ref{fig:cage_bireg_k_4_l_4_connect_initial_layer_3}.

Now we consider connecting the other outgoing edges of each
$\hat{x}_{m,i}$ vertex to the remaining $\hat{y}_{j,\mu}$ vertices, $j
\ne 0$.  The set of mutually-orthogonal squares of order $q = 3$,
given in Example~\ref{ex:mut_orth_squares_q_3}, guarantees that
$4$-cycles do not get introduced in
step~\ref{it:cage_k_q+1_l_q+1_connect_latin_squares}.
Figure~\ref{fig:cage_bireg_k_4_l_4_final} shows the resulting
bipartite cage graph.

\subsection{Properties of Graph Constructed from Algorithm~\ref{alg:cage_k_q+1_l_q+1}}
\label{subsec:reg_cages_properties}

We show that the graph constructed from
Algorithm~\ref{alg:cage_k_q+1_l_q+1} is indeed a cage graph, as well
as discuss additional properties.

\begin{lemma}
\label{thm:alg_cage_k_q+1_l_q+1_has_girth_6}
In the bipartite graph constructed from
Algorithm~\ref{alg:cage_k_q+1_l_q+1}, the shortest cycle consists of
at least $6$ vertices.
\end{lemma}
\vspace{-1.0ex}
\begin{IEEEproof}
See Appendix~\ref{app:proof_thm_alg_cage_k_q+1_l_q+1_has_girth_6} or
\cite[Section~4]{wong:cages}.
\end{IEEEproof}

\begin{theorem}[{see also \cite[Section~4]{wong:cages}}]
\label{thm:alg_cage_k_q+1_l_q+1_is_cage}
The regular bipartite graph constructed from
Algorithm~\ref{alg:cage_k_q+1_l_q+1} is a bipartite cage graph of
girth at least $6$, with degree $q+1$ at all vertices.
\end{theorem}
\vspace{-1.0ex}
\begin{IEEEproof}
Algorithm~\ref{alg:cage_k_q+1_l_q+1} results in $1 + l(k-1) = q^2 + q
+ 1$ vertices for $Y$ and $l + l(l-1)(k-1)/k = q^2 + q + 1$ vertices
for $X$---where every vertex has degree $q+1$.  Thus $v = |Y|$ and $u
= |X|$ achieve the lower bounds of
Lemma~\ref{thm:lower_bounds_num_vertices} for the required degree
distributions.  By Lemma~\ref{thm:alg_cage_k_q+1_l_q+1_has_girth_6},
the shortest cycle has at least $6$ vertices, so the result is shown.
\end{IEEEproof}

By interpreting $Y$ as the elements and $X$ as the blocks, we have
constructed a $S(2,k,v) = S(2,q+1,q^2+q+1)$ Steiner system---and also
a corresponding storage system design.

We see that in order to generate the cage graph and associated block
system, the only required information is the generator element used to
generate the multiplicative group for the finite field---as the set of
mutually-orthogonal squares can then be uniquely determined.  Thus
lookup tables for the entire block design need not be stored, since
the tables can always be generated easily.

\begin{figure}[htbp]
\centering
\subfloat[Bipartite graph.]{
\psfrag{y00}[l][l]{\footnotesize $0$}
\psfrag{y01}[l][l]{\footnotesize $1$}
\psfrag{y02}[l][l]{\footnotesize $2$}
\psfrag{y03}[l][l]{\footnotesize $3$}
\psfrag{y04}[l][l]{\footnotesize $6$}
\psfrag{y05}[l][l]{\footnotesize $5$}
\psfrag{y06}[l][l]{\footnotesize $4$}
\psfrag{x00}[l][l]{\footnotesize $\{0,1,2\}$}
\psfrag{x01}[l][l]{\footnotesize $\{0,3,6\}$}
\psfrag{x02}[l][l]{\footnotesize $\{0,5,4\}$}
\psfrag{x03}[c][l]{\footnotesize $\{1,3,5\}$}
\psfrag{x04}[c][l]{\footnotesize $\{1,6,4\}$}
\psfrag{x05}[c][l]{\footnotesize $\{2,3,4\}$}
\psfrag{x06}[c][l]{\footnotesize $\{2,6,5\}$}
\includegraphics[height=1.15in,scale=0.5]{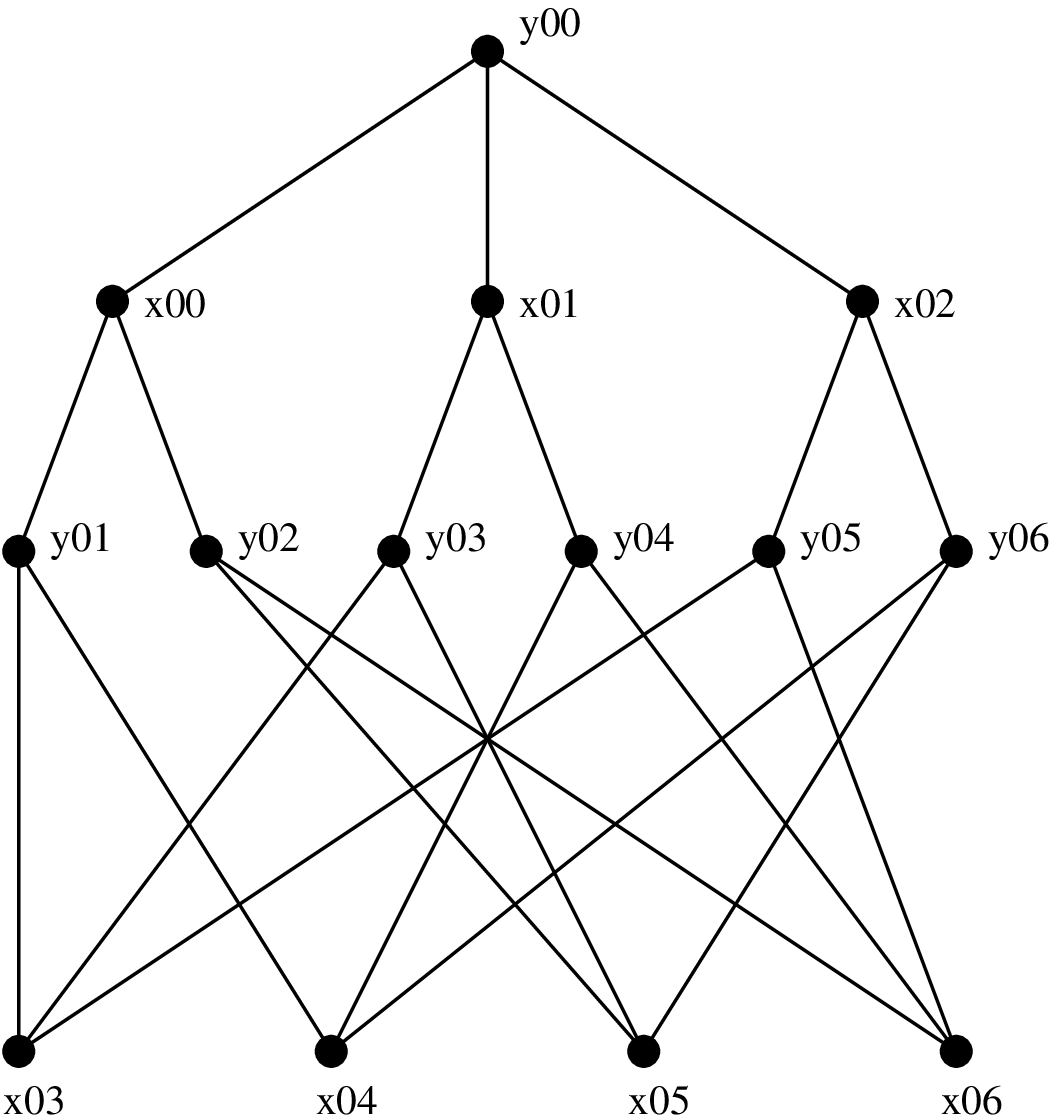}
\label{fig:steiner_sys_cage_bireg_k_3_l_3}
}
\hfil
\subfloat[Block design.]{
\newcommand{\bsz}{\small}  
\newcommand{\gsz}{\small}  
\psfrag{b0}[c][c]{\bsz $b_{0}$}  \psfrag{b1}[c][c]{\bsz $b_{1}$}
\psfrag{b2}[c][c]{\bsz $b_{2}$}  \psfrag{b3}[c][c]{\bsz $b_{3}$}
\psfrag{b4}[c][c]{\bsz $b_{4}$}  \psfrag{b5}[c][c]{\bsz $b_{5}$}
\psfrag{b6}[c][c]{\bsz $b_{6}$}
\psfrag{g00}[c][c]{\gsz $0$}  \psfrag{g01}[c][c]{\gsz $1$}  \psfrag{g02}[c][c]{\gsz $2$}
\psfrag{g10}[c][c]{\gsz $0$}  \psfrag{g11}[c][c]{\gsz $3$}  \psfrag{g12}[c][c]{\gsz $6$}
\psfrag{g20}[c][c]{\gsz $0$}  \psfrag{g21}[c][c]{\gsz $4$}  \psfrag{g22}[c][c]{\gsz $5$}
\psfrag{g30}[c][c]{\gsz $1$}  \psfrag{g31}[c][c]{\gsz $3$}  \psfrag{g32}[c][c]{\gsz $5$}
\psfrag{g40}[c][c]{\gsz $1$}  \psfrag{g41}[c][c]{\gsz $4$}  \psfrag{g42}[c][c]{\gsz $6$}
\psfrag{g50}[c][c]{\gsz $2$}  \psfrag{g51}[c][c]{\gsz $3$}  \psfrag{g52}[c][c]{\gsz $4$}
\psfrag{g60}[c][c]{\gsz $2$}  \psfrag{g61}[c][c]{\gsz $5$}  \psfrag{g62}[c][c]{\gsz $6$}
\includegraphics[height=1.1in,scale=0.5]{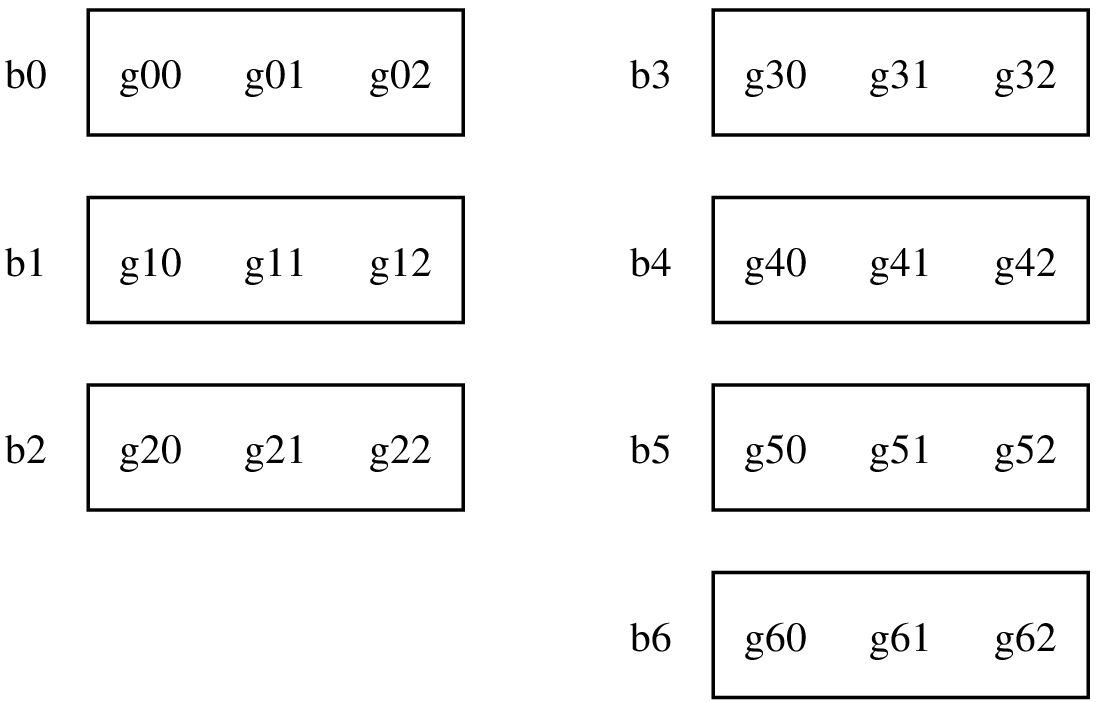}
\label{fig:steiner_3_7}
}
\caption{Steiner system corresponding to $k = 3$, $l = 3$.
Figure~\ref{fig:steiner_sys_cage_bireg_k_3_l_3} visualizes the system
as a bipartite graph, and Figure~\ref{fig:steiner_3_7} shows the block
design.  This gives the same Steiner system as in \cite[Fig.~3
(Example~2)]{elrouayheb:frac_rep_codes}.}
\label{fig:steiner_sys_k_3_l_3}
\end{figure}

In fact, the constructibility of a regular cage graph with $q^2+q+1$
vertices in each vertex set is equivalent to the constructibility of a
projective plane of order $q+1$ \cite{bondy:graph_theory}.  The
regular cage graph with $k = l = 3$ is the Heawood graph; see
Figure~\ref{fig:steiner_sys_k_3_l_3}, which also shows the associated
Steiner system.  This construction of the Heawood graph is analogous
to the Skolem construction~\cite{lindner:design_theory} of Steiner
triple systems for $v = 9$.

We also mention that similar methods can be used to construct regular
graphs (i.e., $k = l = q+1$) of girth~$6$ when $q$ is not a prime
power (e.g., see \cite{okeefe:smallest_graph_girth_6_valency_7}, where
$q = 6$).

\section{Scalable Designs}
\label{sec:scal_designs}

In Section~\ref{subsec:scal_designs_k_q+1_l_pn}, we construct cage
graphs where the vertex degrees of the two vertex sets are highly
unbalanced, i.e., where $\deg(X) = k = q+1$ but $\deg(Y) = l =
p_n(q)$.  We discuss some favorable scalability properties in
Section~\ref{subsec:scal_designs_advantages}.

\subsection{Construction of Designs with $k = q+1$, $l = p_n(q)$}
\label{subsec:scal_designs_k_q+1_l_pn}

\begin{figure}[htbp]
\centering
\psfrag{y00}[l][l]{\footnotesize $y_{0}$}
\psfrag{y01}[l][l]{\footnotesize $y_{1}$}
\psfrag{y02}[l][l]{\footnotesize $y_{2}$}
\psfrag{y03}[l][l]{\footnotesize $y_{3}$}
\psfrag{y04}[l][l]{} 
\psfrag{y05}[l][l]{} 
\psfrag{y06}[l][l]{} 
\psfrag{y07}[l][l]{} 
\psfrag{y08}[l][l]{} 
\psfrag{y09}[l][l]{} 
\psfrag{y10}[l][l]{} 
\psfrag{y11}[l][l]{} 
\psfrag{y12}[l][l]{} 
\psfrag{y13}[l][l]{} 
\psfrag{y14}[l][l]{} 
\psfrag{y15}[l][l]{} 
\psfrag{y16}[l][l]{} 
\psfrag{y17}[l][l]{} 
\psfrag{y18}[l][l]{} 
\psfrag{x00}[l][l]{\footnotesize $x_{0}$}
\psfrag{x01}[l][l]{\footnotesize $x_{1}$}
\psfrag{x02}[l][l]{\footnotesize $x_{2}$}
\psfrag{x03}[l][l]{} 
\psfrag{x04}[l][l]{} 
\psfrag{x05}[l][l]{} 
\psfrag{x06}[l][l]{\footnotesize $x_{l}$}
\psfrag{x07}[l][l]{\footnotesize $x_{l+1}$}
\psfrag{x08}[l][l]{\footnotesize $x_{l+2}$}
\psfrag{x09}[l][l]{\footnotesize $x_{l+3}$}
\psfrag{x10}[l][l]{\footnotesize $x_{l+4}$}
\psfrag{x11}[l][l]{\footnotesize $x_{l+5}$}
\psfrag{x12}[l][l]{} 
\psfrag{x13}[l][l]{} 
\psfrag{x14}[l][l]{} 
\psfrag{x15}[l][l]{} 
\psfrag{x16}[l][l]{} 
\psfrag{x17}[l][l]{} 
\psfrag{L}[r][r]{\scriptsize $p_n(q) = l$}
\psfrag{Kminus1}[r][r]{\scriptsize $q = k-1$}
\psfrag{Lminus1}[r][r]{\scriptsize $q p_{n-1}(q) = l-1$}
\psfrag{K}[r][r]{\scriptsize $q+1 = k$}
\psfrag{cdots}[c][c]{\LARGE $\cdots$}
\psfrag{cdits}[c][c]{\scriptsize $\cdots$}
\psfrag{layer0}[l][l]{} 
\psfrag{layer1}[l][l]{} 
\psfrag{layer2}[l][l]{} 
\psfrag{layer3}[l][l]{} 
\includegraphics[height=1.15in,scale=0.5]{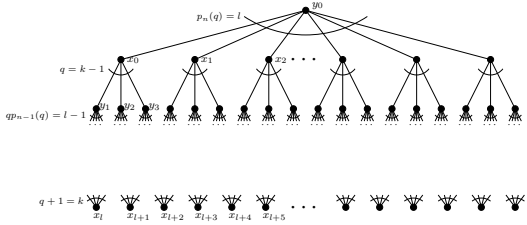}
\caption{Construction of bipartite cage graph with $k = q+1$, $l =
p_n(q)$.}
\label{fig:cage_bireg_k_q1_l_pn}
\end{figure}

The construction here is recursive; thus we call $l[n] = p_n(q)$ as
the degree of the vertices in $Y$, at iteration~$n$.\footnotemark
~(For notational simplicity, if no iteration number is specified, then
it is assumed that we are referring to the quantity for the
$n$\nobreakdash-th iteration.)  The constructed cages have $|X| =
\frac{p_{n+1}(q) p_n(q)}{q+1}$ and $|Y| = p_{n+1}(q)$.  We show such a
graph in Figure~\ref{fig:cage_bireg_k_q1_l_pn}.%
\footnotetext{We let $u[n]$, $v[n]$ denote the respective quantities
at iteration~$n$.  Since $k[n] = q+1$ for all $n$, we do not qualify
$k$ with the iteration number~$n$.}

We will inductively construct bipartite cages with $k = q+1$ and $l[n]
= p_n(q)$ using a layered method similar to before.  Notice that for
$n = 1$, the graph is the $k = l = q+1$ cage.

Thus suppose that a cage graph with parameters $k = q+1$ and $l[n-1] =
p_{n-1}(q)$ exists.  For this graph, $v[n-1] = p_n(q)$ and $u[n-1] =
\frac{p_n(q) p_{n-1}(q)}{q+1}$.  By taking $Y$ as the elements and $X$
as the blocks, this gives a Steiner system with block size $k = q+1$
and with $v[n-1] = p_n(q)$ total elements, i.e., $S(2,q+1,p_n(q))$.
(Here, each element is repeated $l[n-1] = p_{n-1}(q)$ times, and there
are $u[n-1] = \frac{p_n(q) p_{n-1}(q)}{q+1}$ blocks.)  This system can
then be used to construct the $k = q+1$, $l = p_n(q)$ cage---as given
in Algorithm~\ref{alg:cage_k_q+1_l_pn}.

\begin{algorithm}[htb]
\caption{\footnotesize Construction of bipartite cage when $k = q+1$, $l = p_n(q)$}
\label{alg:cage_k_q+1_l_pn}
\footnotesize{
\begin{algorithmic}[1]
\REQUIRE \label{it:cage_k_q+1_l_pn_require_prev_case}
A set of $v[n-1] = p_n(q)$ elements, and a collection $\mathcal{B} =
\{b_h \; | \; h = 0,1,\ldots,u[n-1]-1\}$ of $(q+1)$-element
blocks~$b_h$, such that each element has exactly $l[n-1] = p_{n-1}(q)$
replicas and no particular pair of elements occurs in more than one
block.
\STATE \label{it:cage_k_q+1_l_pn_layer_0}
\textbf{[Layer~$0$]}
Start with a single vertex $y_0 \in Y$.
\STATE \label{it:cage_k_q+1_l_pn_layer_1}
\textbf{[Layer~$1$]}
Connect $y_0$ to $l = p_n(q)$ vertices of $X$.  Without loss of
generality, call these vertices $x_0, x_1, \ldots, x_{p_n(q)-1}$.
\STATE \label{it:cage_k_q+1_l_pn_layer_2}
\textbf{[Layer~$2$]}
For each vertex~$x_j$, $j = 0,1,\ldots,l-1$, connect $x_j$ to $k-1 =
q$ vertices of $Y$.  Let $\hat{y}_{j,m}$, $m = 0,1,\ldots,k-2$, denote
the vertices of this step that are connected to vertex $x_j$.
\FOR{$h=0$ to $u[n-1]-1$}
\STATE \label{it:cage_k_q+1_l_pn_select_subset}
Let the block $b_h$ consist of elements $b_h = \{g_0, g_1, g_2,
\ldots, g_q\}$.
\STATE \label{it:cage_k_q+1_l_pn_layer_3}
\textbf{[Layer~$3$]}
Connect each vertex~$\hat{y}_{g_0,m}$ ($m = 0, 1,\ldots,k-2$) to $q$
distinct vertices of $X$, called $\hat{x}_{m,i}^{(h)}$, $i =
0,1,\ldots,q-1$.  Therefore, $\hat{x}_{m,i}^{(h)} \ne
\hat{x}_{m',i'}^{(h)}$ unless $m = m'$ and $i = i'$.  (For the
$h$\nobreakdash-th iteration, there will be a total of $(k-1)q = q^2$
such vertices $\hat{x}_{m,i}^{(h)}$.)
\STATE \label{it:cage_k_q+1_l_pn_connect_latin_squares}
Consider a vertex~$\hat{x}_{m,i}^{(h)}$, where $m \in
\{0,1,\ldots,k-2\}$, $i \in \{0,1,\ldots,q-1\}$.  Connect
$\hat{x}_{m,i}^{(h)}$ to $\hat{y}_{g_{j+1},L_{i,j}^{(m)}}$, $j =
0,1,\ldots,q-1$.
\ENDFOR
\ENSURE \label{it:cage_k_q+1_l_pn_postcondition}
Bipartite cage with degrees $k = q+1$, $l[n] = p_n(q)$, and number of
vertices $|Y| = v[n] = p_{n+1}(q)$, $|X| = u[n] = \frac{p_{n+1}(q)
p_n(q)}{q+1}$
\end{algorithmic}
}
\end{algorithm}

Algorithm~\ref{alg:cage_k_q+1_l_pn} differs from
Algorithm~\ref{alg:cage_k_q+1_l_q+1} in steps
\ref{it:cage_k_q+1_l_pn_layer_3} and
\ref{it:cage_k_q+1_l_pn_connect_latin_squares} because we only connect
via $\hat{y}_{g_j,m}$ (where $j = 0,1,\ldots,q$) instead of
$\hat{y}_{j,m}$ for all $j = 0,1,\ldots,l-1$.  This is due to only
considering $(q+1)$-element \emph{subsets} instead of the entire set
of $x_0,\ldots,x_{l[n]}$ vertices when constructing each smaller
subcage.

For each iteration~$h$ where we select the subset of layer~$1$
vertices denoted by $b_h = \{g_0, g_1, g_2, \ldots, g_q\}$, let us
call the \emph{$b_h$-subgraph} as the subgraph induced by the subset
of vertices
\begin{eqnarray*}
\lefteqn{ \{y_0\} \cup \{x_j \, | \, j \in b_h\} \cup \{\hat{y}_{j,m}
\, | \, j \in b_h, ~m = 0,1,\ldots,k-2\} }  \nonumber \\
&& \cup \{\hat{x}_{m,i}^{(h)} \, | \, m = 0,1,\ldots,k-2, ~i =
0,1,\ldots,q-1\}
\mbox{.}
\end{eqnarray*}

\begin{lemma}
\label{thm:alg_cage_k_q+1_l_pn_is_valid}
The graph of Algorithm~\ref{alg:cage_k_q+1_l_pn} has the desired
number of vertices, $|X|$ and $|Y|$, and satisfies the degree
requirements.  
\end{lemma}
\vspace{-1.0ex}
\begin{IEEEproof}
This can be shown via careful accounting.  We provide the complete
proof in Appendix~\ref{app:proof_thm_alg_cage_k_q+1_l_pn_is_valid}.
\end{IEEEproof}

\begin{lemma}
\label{thm:alg_cage_k_q+1_l_pn_has_girth_6}
In the constructed bipartite graph of
Algorithm~\ref{alg:cage_k_q+1_l_pn}, the shortest cycle has length of
at least $6$~vertices.
\end{lemma}
\vspace{-1.0ex}
\begin{IEEEproof}
As there are neither odd cycles nor length-$2$ cycles, we only need to
check that there are no length-$4$ cycles.  Since each selection $b_h$
of layer~$1$ vertices induces a subgraph which is isomorphic to the $k
= l = q+1$ bipartite regular cage graph, any properties from the
regular graph also hold for the subgraph.  Thus within any
$b_h$-subgraph, there are no $4$-cycles.

Consequently, any potential $4$-cycle must involve only edges from
layer~$2$ to layer~$3$ vertices, where the layer~$2$ vertices are
connected to different $x_j$ vertices of layer~$1$.  Suppose that the
layer~$2$ vertices $\hat{y}_{j,\mu}$ and $\hat{y}_{j',\mu'}$, where $j
\ne j'$, are involved in a $4$-cycle with the layer~$3$ vertices
$\hat{x}_{m,i}^{(h)}$ and $\hat{x}_{m',i'}^{(h')}$.\footnotemark ~Such
a cycle implies that the $b_h$-subgraph must include the edge between
$\hat{y}_{j,\mu}$ and $\hat{x}_{m,i}^{(h)}$, as well as the edge
between $\hat{y}_{j',\mu'}$ and $\hat{x}_{m,i}^{(h)}$; also, the
$b_{h'}$-subgraph must include the edge between $\hat{y}_{j,\mu}$ and
$\hat{x}_{m',i'}^{(h')}$, as well as the edge between
$\hat{y}_{j',\mu'}$ and $\hat{x}_{m',i'}^{(h')}$.  This means that the
subsets $b_h$ and $b_{h'}$ both contain the elements $j$ and $j'$.
However, since $b_h$ and $b_{h'}$ are two subsets that do not share
any pair of elements, the fact that $j,j' \in b_h$ and $j,j' \in
b_{h'}$ is a contradiction.%
\footnotetext{We know $h \ne h'$, or else the $4$-cycle is entirely
within the $b_h$-subgraph.}
\end{IEEEproof}

\begin{lemma}
\label{thm:alg_cage_k_q+1_l_pn_is_cage}
Supposing that a bipartite cage (of girth~$6$) with parameters $k =
q+1$, $l[n-1] = p_{n-1}(q)$, $v[n-1] = p_n(q)$, and $u[n-1] =
\frac{p_n(q) p_{n-1}(q)}{q+1}$ exists, then
Algorithm~\ref{alg:cage_k_q+1_l_pn} constructs a bipartite cage (of
girth~$6$) with parameters $k = q+1$, $l[n] = p_n(q)$ (and $v[n] =
p_{n+1}(q)$, $u[n] = \frac{p_{n+1}(q) p_n(q)}{q+1}$).
\end{lemma}
\vspace{-1.0ex}
\begin{IEEEproof}
Follows from Lemmas \ref{thm:alg_cage_k_q+1_l_pn_is_valid} and
\ref{thm:alg_cage_k_q+1_l_pn_has_girth_6}.
\end{IEEEproof}

\begin{theorem}
\label{thm:alg_cage_k_q+1_l_pn_constructible_induction}
A bipartite cage of girth~$6$, with parameters $k = q+1$ and $l[n] =
p_n(q)$, exists and is constructible.  This graph has $v[n] =
p_{n+1}(q)$ and $u[n] = \frac{p_{n+1}(q) p_n(q)}{q+1}$.
\end{theorem}
\vspace{-1.0ex}
\begin{IEEEproof}
The base case where $n = 1$ is the $k = l = q+1$ cage graph from
Algorithm~\ref{alg:cage_k_q+1_l_q+1}, and so is constructible.  The
conclusion follows by induction, using
Lemma~\ref{thm:alg_cage_k_q+1_l_pn_is_cage}.
\end{IEEEproof}

\begin{figure}[htbp]
\centering
\newcommand{\bsz}{\small}  
\newcommand{\gsz}{\small}  
\psfrag{b0}[c][c]{\bsz $b_{0}$}  \psfrag{b1}[c][c]{\bsz $b_{1}$}
\psfrag{b2}[c][c]{\bsz $b_{2}$}  \psfrag{b3}[c][c]{\bsz $b_{3}$}
\psfrag{b4}[c][c]{\bsz $b_{4}$}  \psfrag{b5}[c][c]{\bsz $b_{5}$}
\psfrag{b6}[c][c]{\bsz $b_{6}$}  \psfrag{b7}[c][c]{\bsz $b_{7}$}
\psfrag{b8}[c][c]{\bsz $b_{8}$}  \psfrag{b9}[c][c]{\bsz $b_{9}$}
\psfrag{bA}[c][c]{\bsz $b_{10}$} \psfrag{bB}[c][c]{\bsz $b_{11}$}
\psfrag{bC}[c][c]{\bsz $b_{12}$} \psfrag{bD}[c][c]{\bsz $b_{13}$}
\psfrag{bE}[c][c]{\bsz $b_{14}$}
\psfrag{g00}[c][c]{\gsz $0$}  \psfrag{g01}[c][c]{\gsz $1$}  \psfrag{g02}[c][c]{\gsz $2$}  \psfrag{g03}[c][c]{\gsz $7$}
\psfrag{g04}[c][c]{\gsz $8$}  \psfrag{g05}[c][c]{\gsz $9$}  \psfrag{g06}[c][c]{\gsz $10$}
\psfrag{g10}[c][c]{\gsz $0$}  \psfrag{g11}[c][c]{\gsz $3$}  \psfrag{g12}[c][c]{\gsz $6$}  \psfrag{g13}[c][c]{\gsz $11$}
\psfrag{g14}[c][c]{\gsz $14$} \psfrag{g15}[c][c]{\gsz $15$} \psfrag{g16}[c][c]{\gsz $18$}
\psfrag{g20}[c][c]{\gsz $0$}  \psfrag{g21}[c][c]{\gsz $4$}  \psfrag{g22}[c][c]{\gsz $5$}  \psfrag{g23}[c][c]{\gsz $12$}
\psfrag{g24}[c][c]{\gsz $13$} \psfrag{g25}[c][c]{\gsz $16$} \psfrag{g26}[c][c]{\gsz $17$}
\psfrag{g30}[c][c]{\gsz $1$}  \psfrag{g31}[c][c]{\gsz $3$}  \psfrag{g32}[c][c]{\gsz $5$}  \psfrag{g33}[c][c]{\gsz $19$}
\psfrag{g34}[c][c]{\gsz $22$} \psfrag{g35}[c][c]{\gsz $23$} \psfrag{g36}[c][c]{\gsz $26$}
\psfrag{g40}[c][c]{\gsz $1$}  \psfrag{g41}[c][c]{\gsz $4$}  \psfrag{g42}[c][c]{\gsz $6$}  \psfrag{g43}[c][c]{\gsz $20$}
\psfrag{g44}[c][c]{\gsz $21$} \psfrag{g45}[c][c]{\gsz $24$} \psfrag{g46}[c][c]{\gsz $25$}
\psfrag{g50}[c][c]{\gsz $2$}  \psfrag{g51}[c][c]{\gsz $3$}  \psfrag{g52}[c][c]{\gsz $4$}  \psfrag{g53}[c][c]{\gsz $27$}
\psfrag{g54}[c][c]{\gsz $30$} \psfrag{g55}[c][c]{\gsz $31$} \psfrag{g56}[c][c]{\gsz $34$}
\psfrag{g60}[c][c]{\gsz $2$}  \psfrag{g61}[c][c]{\gsz $5$}  \psfrag{g62}[c][c]{\gsz $6$}  \psfrag{g63}[c][c]{\gsz $28$}
\psfrag{g64}[c][c]{\gsz $29$} \psfrag{g65}[c][c]{\gsz $32$} \psfrag{g66}[c][c]{\gsz $33$}
\psfrag{g70}[c][c]{\gsz $7$}  \psfrag{g71}[c][c]{\gsz $11$} \psfrag{g72}[c][c]{\gsz $13$} \psfrag{g73}[c][c]{\gsz $19$}
\psfrag{g74}[c][c]{\gsz $21$} \psfrag{g75}[c][c]{\gsz $27$} \psfrag{g76}[c][c]{\gsz $29$}
\psfrag{g80}[c][c]{\gsz $7$}  \psfrag{g81}[c][c]{\gsz $12$} \psfrag{g82}[c][c]{\gsz $14$} \psfrag{g83}[c][c]{\gsz $20$}
\psfrag{g84}[c][c]{\gsz $22$} \psfrag{g85}[c][c]{\gsz $28$} \psfrag{g86}[c][c]{\gsz $30$}
\psfrag{g90}[c][c]{\gsz $8$}  \psfrag{g91}[c][c]{\gsz $11$} \psfrag{g92}[c][c]{\gsz $12$} \psfrag{g93}[c][c]{\gsz $23$}
\psfrag{g94}[c][c]{\gsz $25$} \psfrag{g95}[c][c]{\gsz $31$} \psfrag{g96}[c][c]{\gsz $33$}
\psfrag{gA0}[c][c]{\gsz $8$}  \psfrag{gA1}[c][c]{\gsz $13$} \psfrag{gA2}[c][c]{\gsz $14$} \psfrag{gA3}[c][c]{\gsz $24$}
\psfrag{gA4}[c][c]{\gsz $26$} \psfrag{gA5}[c][c]{\gsz $32$} \psfrag{gA6}[c][c]{\gsz $34$}
\psfrag{gB0}[c][c]{\gsz $9$}  \psfrag{gB1}[c][c]{\gsz $15$} \psfrag{gB2}[c][c]{\gsz $17$} \psfrag{gB3}[c][c]{\gsz $19$}
\psfrag{gB4}[c][c]{\gsz $20$} \psfrag{gB5}[c][c]{\gsz $31$} \psfrag{gB6}[c][c]{\gsz $32$}
\psfrag{gC0}[c][c]{\gsz $9$}  \psfrag{gC1}[c][c]{\gsz $16$} \psfrag{gC2}[c][c]{\gsz $18$} \psfrag{gC3}[c][c]{\gsz $21$}
\psfrag{gC4}[c][c]{\gsz $22$} \psfrag{gC5}[c][c]{\gsz $33$} \psfrag{gC6}[c][c]{\gsz $34$}
\psfrag{gD0}[c][c]{\gsz $10$} \psfrag{gD1}[c][c]{\gsz $15$} \psfrag{gD2}[c][c]{\gsz $16$} \psfrag{gD3}[c][c]{\gsz $23$}
\psfrag{gD4}[c][c]{\gsz $24$} \psfrag{gD5}[c][c]{\gsz $27$} \psfrag{gD6}[c][c]{\gsz $28$}
\psfrag{gE0}[c][c]{\gsz $10$} \psfrag{gE1}[c][c]{\gsz $17$} \psfrag{gE2}[c][c]{\gsz $18$} \psfrag{gE3}[c][c]{\gsz $25$}
\psfrag{gE4}[c][c]{\gsz $26$} \psfrag{gE5}[c][c]{\gsz $29$} \psfrag{gE6}[c][c]{\gsz $30$}
\includegraphics[height=1.4in,scale=0.5]{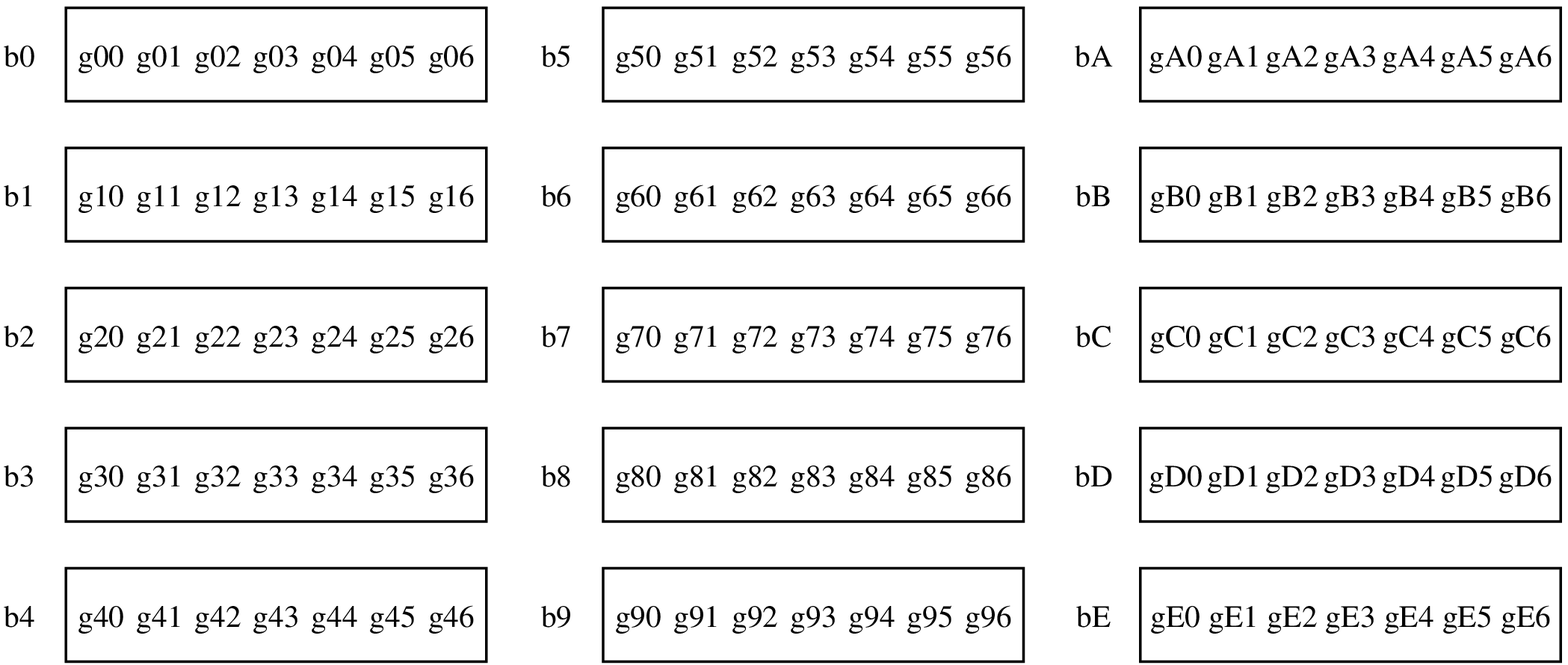}
\caption{Block design for distributed storage system corresponding to
$k = 3$ and $l = 7$.  Each data chunk has $3$~replicas and each
storage node stores $7$~chunks.  In total there are $35$~distinct data
chunks and $15$~storage nodes.}
\label{fig:steiner_3_15_transpose}
\end{figure}

In Figure~\ref{fig:steiner_3_15_transpose}, we show the resulting
storage system design after iteration $n = 2$, for the case $q = 2$
(i.e., $k = 3$).  This system is in fact an extension of the $k = 3$,
$l = 3$ block design of Figure~\ref{fig:steiner_sys_k_3_l_3}; for
storage nodes $b_0, b_1, \ldots, b_6$, the first $3$~data chunks in
each node are exactly the same between Figures \ref{fig:steiner_3_7}
and \ref{fig:steiner_3_15_transpose}.  This scalability will be
explained in Section~\ref{subsec:scal_designs_advantages}.

These cage graphs form a family of designs where $k = q+1$, $l =
p_n(q)$, $v = p_{n+1}(q)$, and $u = \frac{p_{n+1}(q) p_n(q)}{q+1}$,
and thus are coincident with the Steiner systems $S(2, q+1,
p_{n+1}(q))$.

\subsection{Advantages of Scaled Constructions}
\label{subsec:scal_designs_advantages}

The construction of Algorithm~\ref{alg:cage_k_q+1_l_pn} is not merely
a construction for a cage graph with large degree $l[n]$ for the
vertices of $Y$.  This particular construction also allows for the
easy expansion of storage systems built using these methods.  That is,
if an extant system has $l[n-1] = p_{n-1}(q)$, it is relatively simple
to increase the size of the system so that the degree of $Y$ has $l[n]
= p_n(q)$.  This is because the following holds:
\begin{theorem}
\label{thm:scal_alg_cage_k_q+1_l_pn}
Consider a cage graph, $G[n]$, with parameters $k = q+1$, $l[n] =
p_n(q)$ constructed in iteration~$n$ of
Algorithm~\ref{alg:cage_k_q+1_l_pn}.  The cage graph with parameters
$k = q+1$, $l[n-1] = p_{n-1}(q)$ (i.e., constructed in the previous
iteration of Algorithm~\ref{alg:cage_k_q+1_l_pn}, and called $G[n-1]$)
is a subgraph of $G[n]$.
\end{theorem}

Theorem~\ref{thm:scal_alg_cage_k_q+1_l_pn} will be proved with the
help of Lemma~\ref{thm:subset_elements_cage_graph}.
\begin{lemma}
\label{thm:subset_elements_cage_graph}
Consider a cage graph with $k = q+1$, $l[n] = p_n(q)$, to be
constructed in the $n$\nobreakdash-th iteration of
Algorithm~\ref{alg:cage_k_q+1_l_pn}.  From the set of
$p_n(q)$~elements and the collection of blocks~$\mathcal{B}[n]$, it is
possible to select a subset of $p_{n-1}(q)$~elements, called
$\mathcal{S}[n-1]$, such that the subcollection of blocks from
$\mathcal{B}[n]$ that contain only elements from $\mathcal{S}[n-1]$ is
[isomorphic to] the entire collection of blocks $\mathcal{B}[n-1]$
required in the $(n-1)$\nobreakdash-th iteration of the algorithm. 
\end{lemma}
\vspace{-1.0ex}
\begin{IEEEproof}
Here, the elements are $Y$ and the blocks are $X$.  We now prove by
induction.

The base case is $n = 2$.  The cage graph with parameters $k = q+1$,
$l[1] = q+1$ is the graph from Algorithm~\ref{alg:cage_k_q+1_l_q+1}.
In order to construct the cage graph with parameters $k = q+1$, $l[2]
= q^2 + q + 1$ during iteration~$2$, we choose elements from the
collection of blocks $\mathcal{B}[2] = X[1]$ (i.e., the block
collection~$\mathcal{B}$ at iteration~$2$ corresponds to the vertex
set~$X$ at iteration~$1$).  By construction, the block $b_0 \in
\mathcal{B}[2]$ contains $q+1$ elements, so the subgraph associated
with $b_0$ is isomorphic to the cage graph with parameters $k = q+1$,
$l[1] = q+1$.

Now consider an arbitrary iteration~$n$.  From the
$(n-1)$\nobreakdash-th iteration, we know that $\mathcal{B}[n-2]
\subset \mathcal{B}[n-1]$ (up to isomorphism with appropriate indexing
of elements).  Since $\mathcal{B}[n-1]$ is used in iteration~$n$ of
Algorithm~\ref{alg:cage_k_q+1_l_pn} for choosing subsets of layer~$1$
vertices to construct $G[n]$, and $\mathcal{B}[n-2]$ was used in
iteration $n-1$ for choosing subsets of layer~$1$ vertices to
construct $G[n-1]$, then the fact that $\mathcal{B}[n-2] \subset
\mathcal{B}[n-1]$ results in $G[n-1]$ being a subgraph of $G[n]$.  The
block systems corresponding to $G[n-1]$ and $G[n]$ thus satisfy
$\mathcal{B}[n-1] \subset \mathcal{B}[n]$ (again, up to isomorphism
with appropriate indexing).  Because the blocks of $\mathcal{B}[n-1]$
contain a total of $p_{n-1}(q)$ elements (i.e., $\left| \{ y \in b \;
| \; b \in \mathcal{B}[n-1] \} \right| = p_{n-1}(q)$), the result is
shown.
\end{IEEEproof}

Now we can prove Theorem~\ref{thm:scal_alg_cage_k_q+1_l_pn}.
\begin{IEEEproof}[Proof of Theorem~\ref{thm:scal_alg_cage_k_q+1_l_pn}]
From Lemma~\ref{thm:subset_elements_cage_graph}, one can select
$p_{n-1}(q)$ layer~$1$ vertices such that the block system consisting
of only these vertices is isomorphic to $\mathcal{B}[n-1]$.  The
subgraph constructed through these layer~$1$ vertices is thus
isomorphic to the graph of the previous iteration, $G[n-1]$.
\end{IEEEproof}

For the distributed storage system, we take $Y$ as the blocks and $X$
as the elements.  Thus each element has $k = q+1$ repetitions and each
block has size $l = p_n(q)$ (such a system requires a total of $v =
p_{n+1}(q)$ storage nodes and stores a total of $u = \frac{p_{n+1}(q)
p_n(q)}{q+1}$ distinct data chunks).  From
Theorem~\ref{thm:scal_alg_cage_k_q+1_l_pn}, we see that because
$G[n-1]$ is a subgraph of $G[n]$---where the subgraph is a truncation
of outgoing edges from each $Y$ vertex---this means that the blocks of
size $l[n-1] = p_{n-1}(q)$ are truncations of the blocks of size $l[n]
= p_n(q)$.  Equivalently, if we have constructed (using
Algorithm~\ref{alg:cage_k_q+1_l_pn}) the storage system with block
size $l[n-1] = p_{n-1}(q)$, then expanding to a storage system with
block size $l[n] = p_n(q)$ can be accomplished by appending the
remaining outgoing edges from each $Y$ vertex.  No elements need to be
moved from the existing system, and yet the Steiner property (of no
repeating pairs of elements) will still hold---one need only append
new elements to the appropriate blocks.  For instance, the expansion
of the system of Figure~\ref{fig:steiner_3_7} results in the appended
storage system of Figure~\ref{fig:steiner_3_15_transpose}.

It is similarly simple to construct a storage system which has total
number of elements, $\tilde{u}$, that is between the valid quantities
$u[n-1]$ and $u[n]$ (i.e., $u[n-1] < \tilde{u} < u[n]$).  One should
construct the system for $u[n]$~elements (i.e., $k = q+1$ and $l[n] =
p_n(q)$) and then leave empty slots in the blocks which are supposed
to store elements $x_{\tilde{u}}, x_{\tilde{u}+1}, x_{\tilde{u}+2},
\ldots, x_{u[n]-1}$.  This will preserve the Steiner property and also
allow expansion of the storage system until $u[n]$~elements arrive.

\subsection{Other Scalable Constructions}
\label{subsec:scal_designs_other}

Due to space constraints, we do not discuss the construction of a
related class of block designs, which are those that coincide with
affine geometries~\cite{cameron:combinatorics}.  A similar
construction to Algorithm~\ref{alg:cage_k_q+1_l_q+1} can be used to
construct cage graphs where $k = q$ and $l = q+1$---leading to the
graph of Figure~\ref{fig:steiner_sys_cage_bireg_k_3_l_4_bipartite}
when $q = 3$.  From this base case, similar scalability results can be
derived for storage system designs with $k = q$ and $l = p_n(q)$.

\section{Conclusion}
\label{sec:conclusion}

In this paper, we give practical, scalable, and implementable
constructions of bipartite cage graphs where the vertex degrees are
highly asymmetric.  This allows for the design of distributed storage
systems based on Steiner systems, where the number of replicas of each
data chunk may be much smaller than the storage node size.  Using our
constructions, a system designer can guarantee that a system consuming
the least amount of resources (e.g., fewest number of storage nodes)
has been deployed, and also be able to easily expand the storage
system when necessary.

We further comment that the chunk distribution schemes given by our
cage graph construction method can also be used to guarantee collision
resistance in existing storage system implementations.  As an
example, for storage systems implementing distributed hash tables
(DHTs)---such as CAN~\cite{ratnasamy:can}, Chord~\cite{stoica:chord},
Pastry~\cite{rowstron:pastry}, and
Tapestry~\cite{zhao:tapestry}---when the desired replication degree
and number of storage nodes are known, then the chunk and replica
locations from the appropriate block design may be used as the hashing
function.



\bibliographystyle{IEEEtran}
\bibliography{IEEEabrv,references}

\appendices

\section{Latin Squares}
\label{app:latin_squares}

We discuss Latin squares and mutually-orthogonal Latin squares---which
will aid in the construction of bipartite cage graphs of girth~$6$.  A
comprehensive treatment of Latin squares can be found in the text by
D\'{e}nes and Keedwell~\cite{denes:latin_squares}.

\begin{definition}
\label{def:latin_squares}
Consider a $q \times q$ matrix~$L$ where the entries take on values
from $\mathcal{Q} = \{0, 1, 2, \ldots, q-1\}$.  Then $L$ is a
\emph{Latin square} if for every row~$i$, the entries satisfy $L_{i,j}
\ne L_{i,j'}$ whenever $j \ne j'$; and for every column~$j$, the
entries satisfy $L_{i,j} \ne L_{i',j}$ whenever $i \ne i'$.
\end{definition}

\begin{definition}
\label{def:latin_squares_col_natural_order}
A column~$j$ of a square $L$ is considered to be in \emph{natural
order} if the symbols $\{0,1,\ldots,q-1\}$ occur in sequential order,
i.e., $L_{i,j} = i$ for $i = 0,1,\ldots,q-1$.
\end{definition}

In fact---given any Latin square---by labeling symbols and permuting
columns appropriately, we can establish a Latin square with a
specified column in natural order.  Next we define the concept of
orthogonality for Latin squares.

\begin{definition}
\label{def:orth_squares}
A pair of $q \times q$ squares, $L^{(m)}$, $L^{(m')}$, is considered
\emph{orthogonal} if the set of ordered pairs of elements satisfies
$\{ ( L_{i,j}^{(m)}, L_{i,j}^{(m')} ) \; | \; i,j \in \mathcal{Q} \} =
\{(a,b) \; | \; a,b \in \mathcal{Q} \}$.  Thus $L^{(m)}$ and
$L^{(m')}$ are orthogonal if the pairwise catenation of the two
squares takes on all $q^2$ pairs of symbols chosen from $\mathcal{Q}$.
\end{definition}

\begin{definition}
\label{def:mut_latin_squares}
A set of $r$ squares $\{L^{(1)}, L^{(2)}, \ldots, L^{(r)}\}$ (each of
size $q \times q$) is considered to be \emph{mutually-orthogonal} if
every pair of squares, $L^{(m)}$, $L^{(m')}$, where $m,m' =
1,2,\ldots,r$ and $m \ne m'$, are orthogonal.
\end{definition}

When $q$ is a prime or prime power, sets of mutually-orthogonal Latin
squares can be derived by first identifying the generator of the
multiplicative group associated with the finite field of
characteristic~$q$.  That is, consider a Galois field $GF(q)$ with
primitive element~$\alpha$, so that the elements are
\[
e_0 = 0, \;\; e_1 = 1, \;\; e_2 = \alpha, \;\; e_3 = \alpha^2,
\;\; \ldots, \;\; e_{q-1} = \alpha^{q-2}
\mbox{.}
\]
Then the Latin squares, $L^{(1)}, L^{(2)}, \ldots, L^{(q-1)}$, with
entries
\[
L_{i,j}^{(m)} = e_i + e_m e_j, \;\; \forall m = 1,2,\ldots,q-1, ~i,j = 0,1,\ldots,q-1
\]
are mutually-orthogonal with natural order zeroth
column.\footnotemark%
\footnotetext{If $e_i \ne i$, then we can always reorder the rows of
$L^{(m)}$ so that the zeroth column consists of the symbols
$\{0,1,\ldots,q-1\}$ in sequential order.}

If we let the $q \times q$ matrix $L^{(0)}$ consist of $L_{i,j}^{(0)}
= i$ for all $i,j \in \{0,1,\ldots,q-1\}$ (i.e., each column of
$L^{(0)}$ is the same, and consists of symbols numbered sequentially),
then the set of squares $\{L^{(0)}, L^{(1)}, L^{(2)}, \ldots,
L^{(q-1)}\}$ is a set of $q$ mutually-orthogonal squares---where only
$L^{(0)}$ is not Latin.

\begin{lemma}
\label{thm:mut_orth_squares_only_zeroth_col_equal}
For $\{L^{(0)}, \ldots, L^{(q-1)}\}$ as defined above, we have
$L_{i,j}^{(m)} = L_{i,j}^{(m')}$ if and only if $j = 0$.  That is, for
any pair of squares, only the zeroth column has overlapping entries.
\end{lemma}
\vspace{-1.0ex}
\begin{IEEEproof}
Because all of the squares have the zeroth column in natural order,
sufficiency is immediate.  Now, since there are $q$ entries in the
zeroth column, and there are only $q$ pairs of elements $(a,b)$ such
that $a = b$ (where $a,b \in \mathcal{Q}$), by the definition of
mutually-orthogonal squares, we know that no other [non-zeroth] column
will have overlapping entries.
\end{IEEEproof}

\section{Proofs of Selected Lemmas}
\label{app:proofs}

\subsection{Proof of Lemma~\ref{thm:lower_bounds_num_vertices}}
\label{app:proof_thm_lower_bounds_num_vertices}

The lower bound on $v$ can be seen by considering an arbitrary vertex
$y \in Y$.  The vertex $y$ must be connected to $l$ distinct vertices
of $X$; call this subset of vertices $\tilde{X} \subseteq X$.  Now
suppose that two vertices $x_1, x_2 \in \tilde{X}$ were also both
connected to some other vertex $\tilde{y} \ne y$.  Then the graph
would have a cycle of length $4$, consisting of vertices $y \connects
x_1 \connects \tilde{y} \connects x_2 \connects y$.  Thus each vertex
in $\tilde{X}$ must be connected to $k-1$ unique vertices of $Y$; we
let these vertices be $\tilde{Y}$, where $|\tilde{Y}| = l(k-1)$.
Because $|\{y\} \cup \tilde{Y}| = 1 + l(k-1)$ (since $y \not\in
\tilde{Y}$), we establish the lower bound on $v = |Y|$.

Now consider the $l(k-1)$ vertices of $\tilde{Y}$.  These vertices
must each be connected to only one vertex of $\tilde{X}$.  Otherwise,
a vertex $\tilde{y} \in \tilde{Y}$ connected to both $x_1 \in
\tilde{X}$ and $x_2 \in \tilde{X}$ would form the $4$-cycle $\tilde{y}
\connects x_1 \connects y \connects x_2 \connects \tilde{y}$ (similar
to above).  Therefore for any $\tilde{y} \in \tilde{Y}$, the vertex
must connect to at least $l-1$ vertices of $X \setminus \tilde{X}$.
Let $\hat{X}$ consist of vertices in $X \setminus \tilde{X}$ such that
all $x \in \hat{X}$ are connected to some vertex in $\tilde{Y}$.
Since there are at least $l(k-1)(l-1)$ edges between $\tilde{Y}$ and
$\hat{X}$, and any vertex $x \in \hat{X}$ has degree $k$, then
$|\hat{X}| \geq l(k-1)(l-1)/k$.  As $\tilde{X} \cap \hat{X} =
\emptyset$, so $u = |X| \geq |\tilde{X}| + |\hat{X}| \geq l +
l(l-1)(k-1)/k$.

\subsection{Proof of Lemma~\ref{thm:alg_cage_k_q+1_l_q+1_has_girth_6}}
\label{app:proof_thm_alg_cage_k_q+1_l_q+1_has_girth_6}

We show that there are no cycles of lengths $2$ or $4$ (since
bipartite graphs have no odd cycles).  Clearly, there are no
$2$-cycles, since the graph is simple (i.e., no multiple edges).
 
To show that there are no cycles of length~$4$, we consider vertices
from each particular layer, and show that the construction results in
no $4$-cycle involving the vertices at that layer.  For layer~$0$,
there are no $4$-cycles which include vertex~$y_0$, as layers $0$,
$1$, and $2$ form a tree of depth~$3$.  Now consider any $4$-cycles
which include some vertex $x_j$ from layer~$1$.  Such a $4$-cycle must
also include $\hat{y}_{j,m}$ and $\hat{y}_{j,m'}$ for some $m \ne m'$
(and $m,m' \in \{0,1,\ldots,k-2\}$).  If $j = 0$, then
step~\ref{it:cage_k_q+1_l_q+1_layer_3} of the algorithm guarantees
that $\hat{y}_{j,m}$ and $\hat{y}_{j,m'}$ do not connect to any
layer~$3$ vertices in common.  For $j \ne 0$, since any layer~$3$
vertex $\hat{x}_{m,i}$ is connected to at most one vertex of $\{
\hat{y}_{j,\mu} \; | \; \mu = 0,1,\ldots,k-2 \}$, so the vertices
$\hat{y}_{j,m}$ and $\hat{y}_{j,m'}$ can not be connected to the same
layer~$3$ vertex for $m \ne m'$.

This leaves $4$-cycles consisting only of layer~$2$ and layer~$3$
vertices.  Suppose that a vertex $\hat{y}_{0,m}$ is a member of a
$4$-cycle (for any $m \in \{0,1,\ldots,k-2\}$).  Note that
$\hat{y}_{0,m}$ is only connected to the $l-1$ vertices
$\hat{x}_{m,i}$, $i = 0,1,\ldots,l-2$.  Because $L^{(m)}$ has Latin
columns (even for $L^{(0)}$), we see that the layer~$3$ vertices
$\hat{x}_{m,i}$ and $\hat{x}_{m,i'}$, where $i \ne i'$, will never
connect to the same layer~$2$ vertex, i.e.,
$\hat{y}_{j+1,L_{i,j}^{(m)}} \ne \hat{y}_{j+1,L_{i',j}^{(m)}}$ for any
$j = 0,1,\ldots,l-2$.  (Of course, $\hat{x}_{m,i}$ and
$\hat{x}_{m,i'}$ are both connected to $\hat{y}_{0,m}$, but they are
connected to no other common vertex.) Thus, $\hat{y}_{0,m}$ is not
part of a $4$-cycle.

Now consider a potential $4$-cycle consisting of vertices
$\hat{y}_{j,\mu}$ and $\hat{y}_{j',\mu'}$, where $j,j' \ne 0$ and $j
\ne j'$.  Then there will be two layer~$3$ vertices $\hat{x}_{m,i}$
and $\hat{x}_{m',i'}$ such that $L_{i,j-1}^{(m)} = \mu =
L_{i',j-1}^{(m')}$ and $L_{i,j'-1}^{(m)} = \mu' = L_{i',j'-1}^{(m')}$.
However, this would imply that the two squares $L^{(m)}$ and
$L^{(m')}$ have two separate columns, $j-1$ and $j'-1$, where
overlapping entries between the two squares can be found; this
contradicts Lemma~\ref{thm:mut_orth_squares_only_zeroth_col_equal},
since only the zeroth column has overlapping entries.  Thus no
$4$-cycles exist which involve layer~$2$ vertices.

Since layer~$3$ vertices must connect to layer~$2$ vertices, this
implies that the shortest cycle consists of at least $6$ vertices.

\subsection{Proof of Lemma~\ref{thm:alg_cage_k_q+1_l_pn_is_valid}}
\label{app:proof_thm_alg_cage_k_q+1_l_pn_is_valid}

We want to show that all the vertices in $Y$ have exactly $l[n] =
p_n(q)$ outgoing edges, and all the vertices in $X$ have exactly $k =
q+1$ outgoing edges.  Furthermore, $|Y| = v[n] = p_{n+1}(q)$ and $|X|
= u[n] = \frac{p_{n+1}(q) p_n(q)}{q+1}$.

First we verify that we have the correct number of vertices.  For $Y$,
there is $1$ layer~$0$ vertex.  In layer~$2$, we will have $l(k-1) =
p_n(q) q = p_{n+1}(q) - 1$ vertices, since each of the $l$ vertices
$x_j$, $j = 0,1,\ldots,l-1$, is connected to $k-1$ different layer~$2$
vertices of $Y$.  Thus $v[n] = |Y| = p_{n+1}(q)$.  For $X$, there are
$l[n] = p_n(q)$ layer~$1$ vertices.  For layer~$3$, in each of the
$u[n-1]$ iterations of step~\ref{it:cage_k_q+1_l_pn_layer_3}, there
are $(k-1)q = q^2$ distinct vertices of $X$ involved.  Thus, layer~$3$
consists of $q^2 u[n-1] = \frac{q^2 p_n(q) p_{n-1}(q)}{q+1}$ vertices.
Therefore, $u[n] = |X| = p_n(q) + \frac{q^2 p_n(q)p_{n-1}(q)}{q+1} =
\frac{p_{n+1}(q) p_n(q)}{q+1}$.

Now we count the number of edges from each vertex.  For layer~$0$,
step~\ref{it:cage_k_q+1_l_pn_layer_1} results in degree of $l[n] =
p_n(q)$ for vertex~$y_0$.  For layer~$1$, each vertex $x_j$, $j =
0,1,\ldots,l-1$, is connected to exactly $q+1$ vertices (one edge to
$y_0$ and then $q$ edges to the layer~$2$ vertices), as can be seen
from step~\ref{it:cage_k_q+1_l_pn_layer_2}.

Now consider a particular layer~$2$ vertex $\hat{y}_{j,m}$.  We know
in the collection of subsets, $\mathcal{B}$, that each element $j$ is
selected exactly $l[n-1] = p_{n-1}(q)$ times; thus, $\hat{y}_{j,m}$
occurs in exactly $l[n-1] = p_{n-1}(q)$ iterations.  Moreover, in each
iteration that $\hat{y}_{j,m}$ occurs, it has exactly $q$ edges to the
layer~$3$ vertices (whether or not $j$ is the $g_0$ or some $g_{j'+1}$
of the current subset~$b_h$).  Therefore, each layer~$2$ vertex has
$1$ edge to its corresponding layer~$1$ vertex, and $q p_{n-1}(q)$
edges to the layer~$3$ vertices, for a total of $1 + q p_{n-1}(q) =
p_n(q)$ edges---which is the desired degree for that vertex.

By construction, every layer~$3$ vertex $\hat{x}_{m,i}^{(h)}$ has
degree~$q+1$, that is, $1$ edge from the associated $\hat{y}_{g_0,m}$
and $q$ edges to the vertices $\hat{y}_{g_{j+1}, L_{i,j}^{(m)}}$, $j =
0,1,\ldots,q-1$, connected via the Latin squares method.

\end{document}